\documentclass[12pt]{spieman}  
\usepackage{amsmath,amsfonts,amssymb}
\usepackage{graphicx}
\usepackage{setspace}
\usepackage{tocloft}
\usepackage{lineno}
\usepackage{ulem}

\usepackage{fancyhdr}
\title{Tight-binding photonics}

\author[a$\dagger$]{Jing Li}
\author[a$\dagger$]{Aodong Li}
\author[a]{Yutao Chen}
\author[a]{Tao Xiao}
\author[b]{Renwen Huang}
\author[a]{Xiaolu Zhuo}
\author[a]{Jun Guan}
\author[c*]{Zhen Gao}
\author[b*]{Peng Zhan}
\author[b*]{Minghui Lu}
\author[a*]{Biye Xie}
\affil[a]{School of Science and Engineering, The Chinese University of Hong Kong, Shenzhen, Guangdong, 518172, P. R. China.}
\affil[b]{National Laboratory of Solid State Microstructures, Collaborative Innovation Center of Advanced Microstructures, Nanjing University, Nanjing, 210093, China.}
\affil[c]{State Key Laboratory of Optical Fiber and Cable Manufacture Technology, Department of Electronic and Electrical Engineering, Guangdong Key Laboratory of Integrated Optoelectronics Intellisense, Southern University of Science and Technology, Shenzhen, 518055, China.}

\cftpagenumbersoff{figure}
\cftpagenumbersoff{table} 
\begin{document} 
	\maketitle
	
	\begin{abstract}
    Photonics, dealing with the generation, manipulation and detection of photons in various systems, lays the
    foundation of many advanced technologies. A key task of photonics is to know how photons propagate in complex media such as periodic and aperiodic photonic crystals. The conventional wisdom is to numerically solve the Maxwell equations either by dedicated numerical techniques or brute-force finite-element calculations. Recently, the strict analogy between photonic crystals and theoretical tight-binding models provides an unprecedentedly convenient way of understanding the spectra and wavefunctions of photonic systems by mapping the complicated differential equations into matrixed Hamiltonians that can be easily solved through the band theory and exact diagonalization. In this paper, we present a timely review of tight-binding photonics in various platforms, covering fundamental theories, experimental realizations, unique physical effects, and their potential applications. We also provide a brief outlook on the future trends of this active area. Our review offers an in-depth and comprehensive picture on this rapidly developing field and may shed light on the future design on advanced tight-binding photonic devices.
	\end{abstract}
	
	
	{\noindent \footnotesize\textbf{*}Zhen Gao,  \linkable{gaoz@sustech.edu.cn} }
    {\noindent \footnotesize\textbf{*}Peng Zhan,  \linkable{zhanpeng@nju.edu.cn} }
    {\noindent \footnotesize\textbf{*}Minghui Lu,  \linkable{luminghui@nju.edu.cn} }
	{\noindent \footnotesize\textbf{*}Biye Xie,  \linkable{xiebiye@cuhk.edu.cn} }
	
	\begin{spacing}{2}   
		
		\section{Introduction}        
		Photonics empowers advanced technologies in modern industry. For example, light detection and ranging (LiDAR) technologies significantly increase the resolution and response speed of detection in electric cars and therefore provides the possibility of achieving autonomous driving with high safety\cite{li2020lidar}. Besides, a microwave photonic chip has achieved ultrafast analog computation including temporal integration and differentiation with sampling rates up to 256 giga samples per second, significantly outperforming conventional electronic processors\cite{feng2024integrated}. When moving to the terahertz-frequency, photonic waveguide has been shown to enable the improvement of information-transmission efficiency in wireless and on-chip communications\cite{yang2020terahertz}.
		Despite fast development of photonic technology, persistent fundamental challenges arises from notable loss, bandwidth limitation, low transmission efficiency and backscattering of waves that inevitably appear in photonic devices and hence put constraints in advancement of photonic technologies. On the other hand, universal and powerful photonic devices require the integration of multi-functionalities into a single and compact design. These challenges and requirements demand a ceaseless effort in achieving  diverse and robust manipulation of photons in complicated structures.
		
		One of the milestones in the manipulation of photons in complex dielectric media occurred in 1987, when E. Yablonovitch and S. John\cite{yablonovitch1987inhibited,john1987strong} clearly put forward the concept of photonic crystals (PCs). These are periodically modulated dielectric materials that can lead to multiple-Bragg diffraction and destructive interference of light, which results in the formation of photonic bandgaps within specific frequency ranges where light propagation is forbidden. Besides the intriguing photonic bandgaps, PCs also give unprecedented ability in controlling the spatial distributions and dispersion of light. Consequently, PCs have been widely utilized in various advanced photonics technologies, such as robust terahertz transmission through sharply bent silicon waveguides\cite{yang2020terahertz}, as well as high-power, single-mode surface-emitting lasers based on open-Dirac electromagnetic cavities\cite{contractor2022scalable} and Dirac-vortex topological cavities\cite{yang2022laser}.   
        
        However, conventional PC designs rely on directly tackling the Maxwell's equations\cite{joannopoulos2008molding}, either through sophisticated theoretical techniques, such as plane wave expansion method\cite{shi2004plane}, $k \cdot p$ theory\cite{peter2010fundamentals}, or by using direct numerical solution methods such as finite element method, and finite-difference time-domain method\cite{lourtioz2005photonic}. While the former method is case-dependent and have very limited applicability range, the brute-force calculation typically requires large amount of calculation resources, especially in 3-dimensional (3D) cases with complex structure. For instance, a finite-element calculation of 3D dielectric PC with 5 $\times$ 5 $\times$ 5 unit cells in each dimension typically has meshing number that exceeding several million degrees of freedom, and as a result, requires several hundred gigabytes in memory and several days in calculation when run on a normal workstation\cite{lourtioz2005photonic}. Currently the ability to achieve high-throughput design of advanced photonic devices is significantly limited. Finding a way to quickly and rigorously obtain the basic physical properties of PCs without the requirement of huge amounts of computational resources is in high demanded and will  significantly boost the development of photonic technology.
                
        Tight-binding (TB) model, originating from condensed matter physics, describes the motion of electrons that are tightly bound around ions in solids\cite{kittel2018introduction}. When we neglect the electron-electron or electron-phonon interaction, the single-particle Hamiltonian of TB model is simply characterized by a square matrix where the diagonal elements describe the on-site energy of electrons and the off-diagonal elements characterize couplings between electrons at different sites. Despite the simple mathematical descriptions, TB model precisely characterizes the underlying physics of electron motions and predicts the existence of many sophisticated physical effects. Intriguingly, by mimicking the lattice sites by photonic meta-atoms, it is possible to qualitatively design PCs purely based on TB models without tackling the complex Maxwell equations. Such kind of photonic systems is denoted as the TB photonics. Specifically, in TB photonics, one can replace the task of solving of Maxwell’s equations by exact diagonalization of the matrixed Hamiltonian to obtain both the eigenvalues and eigenstates. This approach offers a novel theoretical paradigm for PC design, significantly reducing computational resources, while also providing qualitatively accurate characterization of the underline physics. Besides, based on TB photonics, one can conveniently analyse the symmetries, gauge fields, orbitals and other physical quantities in PCs and acquire an intuitive insight into the wave propagation mechanism in PCs. 
        
        Despite these advantages, there are also challenges in realizing TB photonics. One of the main challenges is the difficulty in precisely mimicking all the aspects of TB models such as the on-site and coupling terms by photonic systems. Another difficulty lies in the fact that the 3D electromagnetic wave is generally a vector field while the wavefunction of a typical TB model is described by a scalar field. Nevertheless, partially and qualitatively mimicking TB model has led to designs of many unique photonic systems, especially in topological photonics\cite{lu2014topological,ozawa2019topological,xie2021higher,lin2023topological}. However, strictly realizing the TB photonics is still a difficult task. It is important to note that not all photonic platforms can be mapped onto TB models; therefore, this review focuses exclusively on photonic systems where TB models can be realized. Several excellent review articles have comprehensively covered different aspects of topological photonics. For example, L. Lu et al. \cite{lu2014topological} and Ozawa et al. \cite{ozawa2019topological} provide overviews of topological photonics, while B. Y. Xie et al. \cite{xie2021higher} focus on higher-order topological phenomena and their realizations in photonic and acoustic systems. More recently, Z. K. Lin et al. \cite{lin2023topological} discuss topological effects associated with defects. In this work, we provide a thorough review of the precise realization of TB models in various photonic systems by simulating various aspects of TB model through photons, discussing physical principles and experimental design methods. We also compare the differences and advantages between different platforms of realizing TB photonics.     
        
        In TB photonics, various aspects can be engineered to manipulate the localization and propagation of photons. First, the magnitude of the on-site term of TB photonics can be tuned by changing the radius/dielectric constant in dielectric rods. Dissipation and enhancement of the on-site energy term can be achieved by introducing lossy and gain materials into the dielectric rods, respectively. Orbital degrees of freedom such as $s$-, $p$-orbitals as shown in Fig.~1a, can be realized by using higher-order resonance modes of the dielectric rods. In addition, breaking of TR symmetry can be achieved by employing gyromagnetic materials under an external magnetic field\cite{lu2014topological,ozawa2019topological,li2023exceptional}. Another important aspect of TB photonics is the coupling term characterizes the coupling of fields between resonating modes, which can be Hermitian (positive, negative, complex) and non-Hermitian (Fig.~1b). The magnitude and sign of the coupling coefficient can be controlled via adjusting the distance between dielectric rods and coupling different resonance modes, respectively. Additionally, non-Hermitian coupling can be further realized by using magneto-optical materials or time modulation\cite{lu2014topological,ozawa2019topological,li2023exceptional}. Moreover, the lattice geometry in TB photonics provides a rich playground for realizing diverse symmetry classes and topological phases. For example, periodic lattices with discrete rotational symmetry, such as $C_{3v}$, $C_{4v}$ and $C_{6v}$, can be constructed using triangular, square, and honeycomb arrangements of dielectric units, respectively (Fig.~1c). Beyond these, more exotic geometries such as Penrose-type quasicrystals\cite{penrose1974role}, fractal lattices\cite{gefen1980critical} can be realized through precise spatial arrangement of photonic atoms. These geometrical variations directly affect the photonic band dispersion, topological invariants, and wave localization properties\cite{lu2014topological,ozawa2019topological}. Besides these local aspects, global effective fields can also be introduced in TB photonics, providing extra way to manipulate photons (see Fig.~1d). For example, a linear potential gradient results in a homogeneous effective magnetic field, which can be achieved through variation in dielectric constants or geometric configuration (e.g., varying the lattice spacing) across the PCs\cite{ozawa2019topological,aidelsburger2018artificial}. Moreover, by introducing a Kekulé distortion into a lattice, one can precisely modulate the lattice structure to create a spatially varying mass term with a vortex-like profile\cite{iadecola2016non}. This induces a vortex lattice state, where the phase of the mass term winds around a core in real space. In TB photonics, such U(1) gauge fields can be realized by precisely shifting the positions of dielectric units to create the desired Kekulé distortion, thereby simulating a synthetic effective mass field\cite{gao2020dirac}. Furthermore, more complex effective fields such as SU(2) gauge field can also be engineered in TB photonics, typically by introducing synthetic spin degrees of freedom and enabling it via dynamically modulated coupling\cite{cheng2023artificial}. Finally, the boundary condition is another key factors in forming exotic photonic states. As illustrated in Fig.~1e, domain wall between two distinct lattices, such as those between topologically trivial and nontrivial domains\cite{xie2018second}, or low dimensional domain walls including disclination and dislocation lattices, can host localized edge states with distinct propagation characteristics\cite{xie2022photonic,lin2023topological}. In particular, even for the same lattice structure, different open boundary conditions (OBCs), such as zigzag and armchair edges in honeycomb lattices, can also exhibit remarkably different transport properties\cite{rechtsman2013topological}. These interfaces can be experimentally realized by spatially aligning dielectric units, and carefully fabricating the domain wall. Such configurations are critical in realizing topological edge transport, higher-order topological modes, and robust signal guiding immune to disorder\cite{lu2014topological,ozawa2019topological}. In summary, the synergy of on-site terms, coupling terms, lattice geometry, effective fields, and interface conditions allows for an exceptionally versatile design space for TB photonics. Consequently, it enables precise tuning of mode frequencies, spatial field profiles, bandgaps, and topological features, laying the foundation for wave design in TB photonics.           
 
        In this review, we focus on the construction principles of TB models with diverse aspects, and their realization in photonic systems. This paper is structured as follows: The first section reviews various 1D, 2D and 3D TB models, including non-Hermitian models, topological Wannier-type models, quantum spin Hall models, and non-periodic models. The second section surveys various 1D and 2D TB models formed by photonic waveguide arrays. The third section discusses the works on implementing TB models using coupled optical resonators. The fourth section investigates the realization of TB models in photonic systems with synthetic dimension. In the fifth section, we specifically highlight recent advancements in confined-Mie resonance PCs (CMR-PCs) for achieving perfectly 2D and 3D photonic TB models. Finally, we conclude with a summary and outlook, pointing out that key future directions include achieving long-range and non-Hermitian coupling, as well as constructing complex 3D TB models to unlock richer physical phenomena and high-performance photonic devices. 
         
        \label{sect:intro}  
        \begin{figure}[H]
         	\setlength{\abovecaptionskip}{-15pt}
\begin{center}
	\begin{tabular}{c}
		\includegraphics[height=14cm]{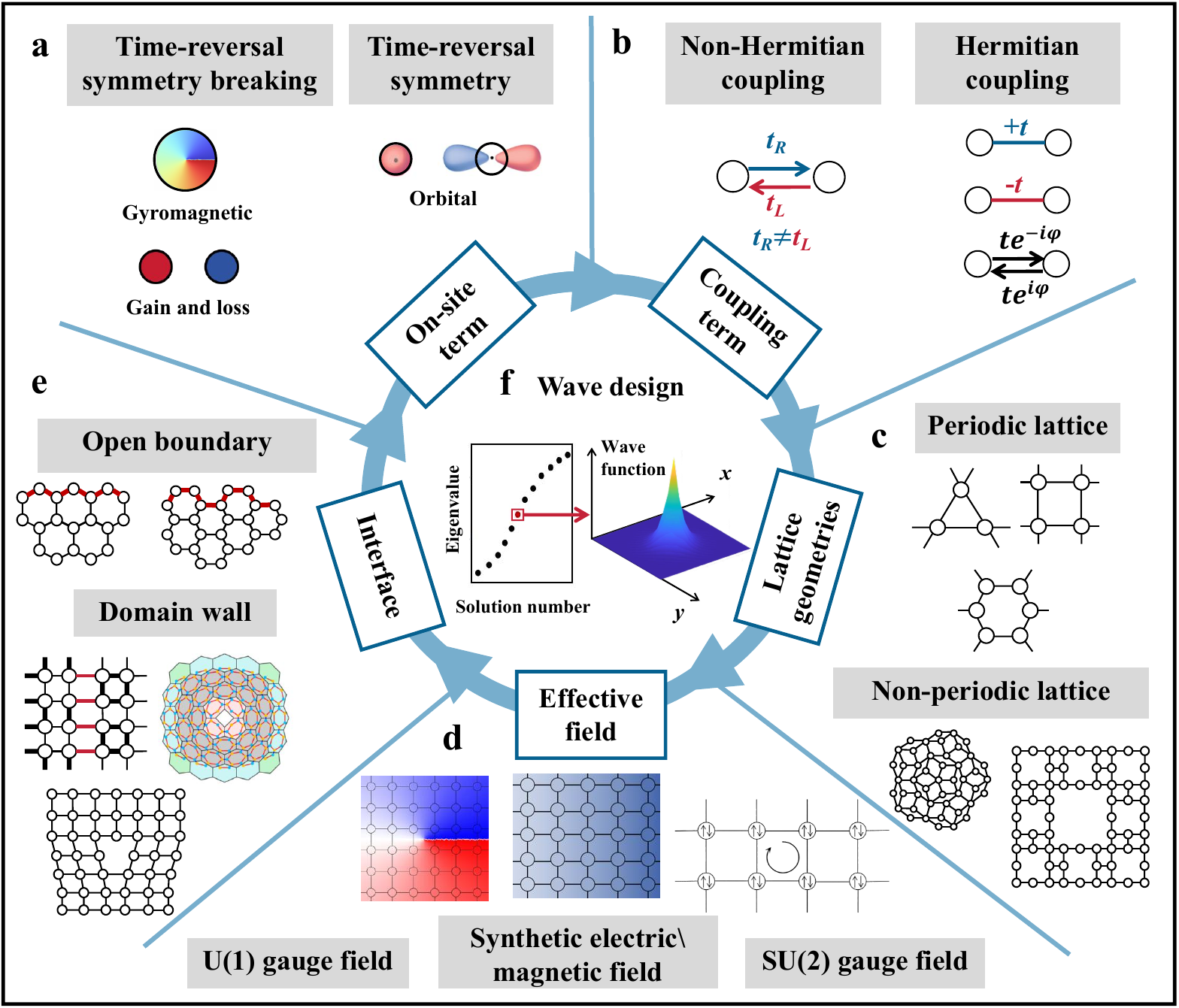}
	\end{tabular}
\end{center}
        	\caption 
        	{Various aspects that control the physical properties of TB models in photonics. \textbf{a}. On-site terms that characterize the intrinsic properties of the lattice site, including TR symmetric and TR symmetry breaking term. \textbf{b}. Coupling terms that describe interactions between sites and can be Hermitian or non-Hermitian. \textbf{c}. Lattice geometries that characterize spatial arrangements of lattice sites, ranging from periodic to non-periodic lattices. \textbf{d}. Effective fields that introduce the global modulation strategies, such as U(1) gauge field, SU(2) gauge field and synthetic electric/magnetic field. \textbf{e}. Interface that characterize the boundary features of lattice, including OBCs and domain walls. \textbf{f}. Physical properties of PCs — including the band spectra and wave functions — are controlled by multiple aspects. } 
        \end{figure}

		\section{Introduction of tight-binding models}		
		\subsection{One-dimensional tight-binding models} 
        The TB model is an approximate method to determine the electronic structure of a system, where the wave functions are constructed as linear combinations of atomic orbital wave functions (LCAO)\cite{harrison2012electronic}. So far, research on 1D TB models has significantly shaped the development of condensed matter physics, especially in the field of topological phases of matter\cite{jackiw1976solitons}. As a prototypical example of 1D TB models: the Su-Schrieffer-Heeger (SSH) model\cite{su1979solitons}, originally proposed by W. P. Su, J. R. Schrieffer, and A. J. Heeger to describe soliton-driven conductivity transitions in polyacetylene, has emerged as a cornerstone for understanding topological phenomena in TB models. The SSH model describes the nearest-neighbor hopping of spinless fermions in a 1D chain with two alternating bonds, as shown in Fig.~2a. Each unit cell contains A and B sublattice sites in the SSH model. The Hamiltonian of the SSH model is given by $\hat{H}_{\text{SSH}} = \sum_{n=1}^{N} \Bigl(v\,\hat{b}_{n}^{\dagger}\,\hat{a}_{n}\;+\; w\,\hat{a}_{n+1}^{\dagger}\,\hat{b}_{n} \;+\;\text{H.c.}\Bigr)$, where $\hat{a}_n$ and $\hat{b}_n$ are the annihilation operators for A and B sublattice sites at position \(n \in \{1, 2, 3, \dots, N\} \), $N$ denotes the total number of unit cells, $v$ and $w$ represent the intracell and intercell hopping amplitudes, respectively. Notably, the band gap of SSH model closes at $v = w$, signaling a topological phase transition characterized by the change of topological invariant (winding number in SSH model). In the topological non-trivial phase ($ w > v $), the SSH model hosts robust zero-energy edge states protected by bulk-boundary correspondence, in which the number of zero-energy states at each edge is equal to the bulk topological invariant, as experimentally verified in quantum simulator platforms\cite{meier2016observation}. Moreover, various photonic applications based on the SSH model have been proposed, such as topological lasing\cite{st2017lasing} and photoluminescence polarization control\cite{tripathi2020topological}. 
        
        While the SSH model primarily highlights the role of coupling terms in exploring the topological matter phase, the Aubry-André-Harper (AAH) model emphasizes the significance of on-site terms, which was first proposed by S. Aubry and G. André in 1980\cite{aubry1980analyticity}. The AAH model illustrates how a periodically varying on-site potential can influence the behavior of quantum particles, as shown in Fig.~2b. The Hamiltonian of the AAH model is $\hat{H}_{\text{AAH}} = \sum_{n=1}^{N} \Bigl(V_n\,\hat{a}_{n}^{\dagger}\,\hat{a}_{n}\;+\; t\,\hat{a}_{n}^{\dagger}\,\hat{a}_{n+1} \;+\; t\,\hat{a}_{n+1}^{\dagger}\,\hat{a}_{n}\Bigr)$, where $V_n = V$cos$(2\pi\beta n+\phi)$ represents the quasiperiodic on-site modulation at site $n$, $V$ is the variation amplitude of the on-site energies, $\phi$ is a relative phase and $\beta$ is the period of the on-site potential modulation in units of the lattice constant. The interaction between the on-site potential and hopping terms dictates the localization behavior of the system. At a critical modulation strength $V = 2t$, the eigenstates of the AAH model undergo a transition from extended to localized\cite{longhi2021phase}. Remarkably, this model paves the way for a range of topologically protected light-emitting devices, including low-cost and energy-efficient integrated laser sources\cite{pilozzi2016topological}.
        
		Beyond the Hermitian examples discussed above, non-Hermitian extensions of TB models have also revealed profound topological phenomena, as exemplified by the Hatano-Nelson (HN) model (Fig.~2c)\cite{hatano1996localization}. The HN model is described by a Hamiltonian that incorporates non-reciprocal hopping terms \(t_R\) $\neq$ \(t_L\): $\hat{H}_{\text{HN}} = \sum_{n=1}^{N} \Bigl(t_L\,\hat{a}_{n}^{\dagger}\,\hat{a}_{n+1}\;+\; t_R\,\hat{a}_{n+1}^{\dagger}\,\hat{a}_{n}\Bigr)$. Moreover, the non-reciprocal hopping gives rise to an exotic feature unique to non-Hermitian systems: a macroscopic number of eigenstates become exponentially localized at a boundary under OBC, which is known as the non-Hermitian skin effect (NHSE)\cite{yao2018edge}. The discoveries of non-Hermitian TB models and skin effect have improved the theoretical framework of the conventional bulk–boundary correspondence\cite{kunst2018biorthogonal}, and opened new possibilities for device applications. For instance, several new topological lasers based on robust non-Hermitian boundary and interface states have been realized\cite{bahari2017nonreciprocal}.   
			
	    \subsection{Two-dimensional tight-binding models}
         Dimensionality is an intrinsic factor that determines many physical properties of systems. The propagation and interaction of wave functions differ dramatically in different dimensions, resulting in fundamentally distinct physical phenomena. Compared to 1D systems, 2D TB models enable access to a richer variety of topological phases and novel physical effects. This subsection introduces numerous 2D TB models, including the graphene lattice\cite{castro2009electronic}, the Haldane model\cite{haldane1988model}, the Kane-Mele model\cite{kane2005quantum}, the Benalcazar-Bernevig-Hughes (BBH) models\cite{benalcazar2017quantized}, the 2D SSH model\cite{xie2018second}, the kagome lattice\cite{ezawa2018higher} and 2D non-periodic TB models\cite{lin2023topological, peterson2021trapped, biesenthal2022fractal, yu2020topological}.
	     
	     Among 2D TB models, the Haldane model holds a place of paramount importance because it is the first to demonstrate that the quantum Hall effect can arise intrinsically from the band structure, without any net magnetic field. The Haldane model is built upon on a honeycomb lattice where atoms are arranged in hexagonal symmetry with two sublattices (black and white circles), as shown in Fig.~2d. In the case of only nearest-neighbor hopping (solid lines), the honeycomb lattice hosts massless Dirac fermions at low energies\cite{castro2009electronic}. Haldane's key novelty was to add next-nearest-neighbor couplings with complex phases (Fig.~2d, dashed lines), which break TR symmetry. Crucially, this is achieved while maintaining zero net magnetic field per hexagonal plaquette via staggered fluxes \cite{haldane1988model}. These complex next-nearest-neighbor couplings open a topologically nontrivial bandgap, giving rise to chiral edge states protected by a nonzero Chern number (or TKNN invariant)\cite{thouless1982quantized}. The Haldane model thus reveals that the realization of the quantum Hall effect requires TR symmetry breaking rather than the presence of a net magnetic field. Since the Haldane model supports dissipationless edge states without the need for an external magnetic field, its experimental realization paves the way for developing low-power-consumption, topological quantum electronics\cite{chang2013experimental}.
	     
	     Unlike the quantum Hall effect, the Kane–Mele model realizes nontrivial topological phases without breaking TR symmetry\cite{kane2005quantum}. It can be viewed as two TR copies of Haldane’s Chern insulator, one  governing spin-up electrons and the other governing spin-down electrons. As illustrated in Fig.~2e, electrons can hop to their next-nearest neighbors with a spin-dependent complex hopping amplitude. The opposing chiralities lead to opposite effective magnetic fields, and preserve TR symmetry. As a result, two counter-propagating helical edge states emerge, where spin-up and spin-down electrons flow clockwise and anticlockwise, respectively.\cite{kane2005quantum,kane2005z} These edge states are protected by the $\mathbb{Z}_2$ topological invariant, which has only two distinguished integer values. Such topological systems are called quantum spin Hall insulators. 
	     
	     Recent advances in higher-order topological insulators (HOTIs) reveal a novel topological paradigm, in which the dimensions of topologically protected edge states are lower than the systems, governed by higher-order bulk-boundary correspondence. The representative Benalcazar-Bernevig-Hughes (BBH) model has firstly predicted the existence of 0D topological corner states in 2D TB model, with its topological phases characterized by quantized quadrupole moments\cite{benalcazar2017quantized}. As shown in Fig.~2f, the BBH model introduces staggered positive (black lines) and negative (red lines) couplings on a square lattice, resulting in \(\pi\)-flux for each plaquette.  Remarkably, incorporating long-range hopping in the 2D BBH model extends the topological phases from $\mathbb{Z}_2$ classified to $\mathbb{Z}$ classified. As the topological invariant increases, multiple 0D corner modes emerge at each lattice corner\cite{benalcazar2022chiral}. Moreover, the \(\pi\)-flux gauge field induces equally profound momentum-space topology reconstruction. For instance, the carefully designed negative couplings fundamentally reconfigure the fundamental domain of the Brillouin zone into a Klein bottle\cite{chen2022brillouin}. This radical transformation originates from the projective symmetry algebra imposed by the \(\pi\)-flux gauge field, which modifies the periodic boundary conditions of Bloch waves.
	     
	     \begin{figure}[H]
	     			      \setlength{\abovecaptionskip}{-15pt}
	     			\begin{center}
	     				\begin{tabular}{c}
	     					\includegraphics[height=10cm]{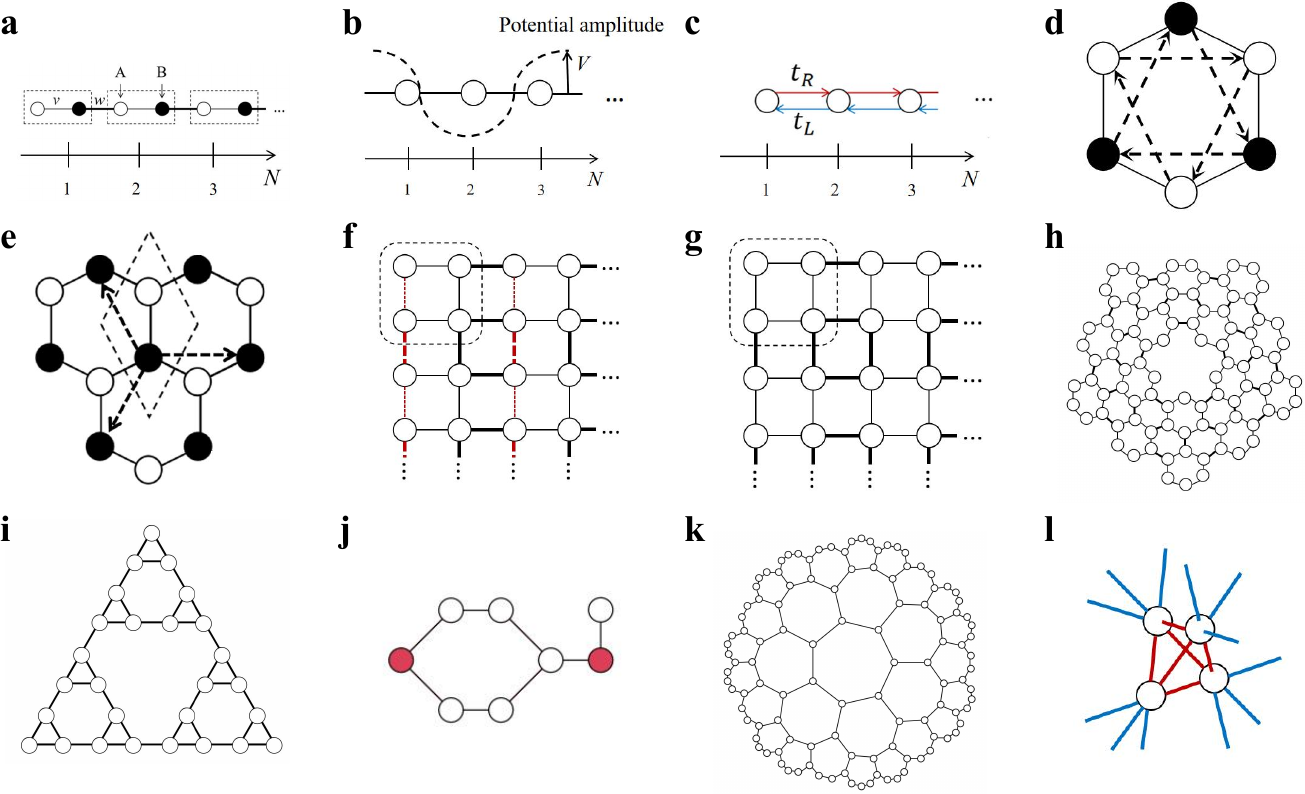}
	     				\end{tabular}
	     			\end{center}
	     	\caption 
	     	{Schematic diagrams of various TB models. \textbf{a}. 1D Su-Schrieffer-Heeger (SSH) chain with  N unit cells, where each unit cell contains two sublattice sites (A and B). Thin lines represent intracell coupling $v$ and thick lines represent intercell coupling $w$. Figure adapted from Ref.~\cite{ozawa2019topological} . \textbf{b}. The Aubry-André-Harper (AAH) model featuring a quasiperiodic potential $V$. Figure adapted from Ref.~\cite{longhi2021phase} . \textbf{c}. Schematic of the Hatano-Nelson (HN) model, illustrating non-reciprocal couplings \(t_R\) (red) and \(t_L\) (blue). Figure adapted from Ref.~\cite{yao2018edge} . \textbf{d}. Schematics for the Haldane model on the honeycomb lattice, indicating complex next-nearest-neighbor couplings (dashed lines). Figure adapted from Ref.~\cite{haldane1988model} . \textbf{e}. The Kane–Mele model is constructed by combining two TR copies of the Haldane model, one for each spin. Figure adapted from Ref.~\cite{kane2005z} . \textbf{f}. Schematics of the 2D Benalcazar-Bernevig-Hughes (BBH) model. Solid black lines and dashed red lines represent positive and negative coupling respectively. Figure adapted from Ref.~\cite{benalcazar2017quantized} . \textbf{g}. The 2D SSH model with positive couplings (solid lines). Figure adapted from Ref.~\cite{xie2018second} . \textbf{h}. A disclination in a hexagonal lattices, formed by cutting off one sector and gluing the remaining parts together. Figure adapted from Ref.~\cite{lin2023topological} . \textbf{i}. Schematics of fractal lattice--Sierpinski gasket. Figure adapted from Ref.~\cite{biesenthal2022fractal} . \textbf{j}. An example of TB models with latently symmetric sites (marked in red). Figure adapted from Ref.~\cite{smith2019hidden} . \textbf{k}. Conformal projection of a hyperbolic heptagon lattice into the Euclidean plane. Figure adapted from Ref.~\cite{kollar2019hyperbolic} . \textbf{l}. Unit cell of breathing pyrochlore lattice. Red and blue lines represent intracell couplings and intercell couplings respectively. Figure adapted from Ref.~\cite{ezawa2018higher} . 
	     	}
	     \end{figure}
	     
	     Despite the theoretical simplicity of these models, the implementation of negative coupling in PCs poses significant experimental challenges. To address this challenge, a substantial number of of recent works have focused on realizing HOTIs with only positive couplings. As an example, B. Y. Xie et al.\cite{xie2018second} propose 2D SSH model with 0D corner states through generalizing 1D SSH model to 2D (see Fig.~2g). In this 2D SSH model, a topologically nontrivial phase can be achieved when intercell couplings are stronger than intracell couplings. These topological properties can be characterized by the dipole polarization as $P= \frac{1}{2\pi} \int_{\text{BZ}} \mathrm{d}k_x \,\mathrm{d}k_y \, \mathrm{Tr}\bigl[ A(k_x, k_y) \bigr]$, where \(A = i\langle \Psi \vert \partial_{k} \vert \Psi \rangle\) is the Berry connection, and the integration is taken over the first Brillouin zone. Hence, this bulk polarization indicates the existence of higher-order topological states in 2D SSH models.\cite{xie2021higher,xie2018second}. Furthermore, topological corner states have also been proposed in other TB models with diverse lattice geometries, include the breathing Kagome lattices\cite{ezawa2018higher,benalcazar2019quantization}, graphdiyne\cite{sheng2019two} and modified Kane-Mele model\cite{ren2020engineering}. These topological corner states can serve as high-Q cavity modes, thus providing potential applications in topological nanolasers\cite{zhang2020low} and high-quality optical hotspots\cite{liu2021high}.  
         
         In the periodic TB models described above, topological invariants are ordinarily defined in momentum-space, relying on Bloch’s theorem and band structure, so that the global band topology is entirely encoded within a single unit cell. However, non-periodic topological insulators can also support nontrivial topological phases and even enable unprecedented physical phenomena. An example is the TB model on a honeycomb lattice with a disclination. Such model is constructed through a cutting-and-gluing process and can host topological disclination states, as illustrated in Fig.~2h\cite{ruegg2013bound}. The disclination defects can robustly trap fractional charges, and the trapped charge can indicate non-trivial crystalline topology.\cite{peterson2021trapped,liu2021bulk} Furthermore, R. Jackiw and P. Rossi predicted a zero mode for the Dirac equation in a vortex background\cite{jackiw1981zero}. The vortex mode can be realized in honeycomb lattice with a Kekulé distortion perturbations\cite{hou2007electron}. In contrast to topological boundary states, topological defects can induce novel physical phenomena and application including unconventional topological pumping via topological defects\cite{xie2022photonic} and Dirac-vortex laser\cite{gao2020dirac}.  
         
         Another noteworthy class of non-periodic TB models is based on fractal lattices, with the Sierpinski triangle\cite{sierpinski1915courbe} serving as a representative example (see Fig.~2i). This self-similar structure arises through an iterative process in which an equilateral triangle is partitioned into four identical segments, and the central region is removed at each iteration. Despite the absence of global translational symmetry, fractal lattices may exhibit other symmetries, such as scale invariance, rotational symmetry, or local point-group symmetries, which play a crucial role in the classification and protection of their topological states. Recent studies have confirmed the persistence of topological edge states in this fractal lattice under an appropriate modulation, such as the periodic driving of the lattice that induces an effective field\cite{yang2020photonic}. Remarkably, the fractal lattices have the potential to accelerate topological protected transport and advance high-end sensing devices\cite{biesenthal2022fractal}. Additionally, there exist other irregular lattice models, such as Penrose lattices and the Hat monotile model, in which localized modes have also been realized through Sutherland loop method\cite{kohmoto1986electronic,schirmann2024physical}. 
         
         Moreover, it is worth emphasizing that graph theory has revealed a multitude of intriguing 2D TB models. Different from typical non-periodic lattices, they exhibit a wealth of interactions among their nodes (vertices) and edges (links)\cite{west2001introduction}. Meanwhile, concepts from graph theory can also provide new perspectives on TB lattices. For example, latent symmetry refers to a hidden symmetry that is not evident in the original Hamiltonian but can be uncovered through subsystem partitioning, leading to an isospectral effective Hamiltonian. This concept helps explain the origin of spectral degeneracies in certain TB models (see Fig.~2j) \cite{smith2019hidden}. Such findings may inspire novel approaches for modifying TB models asymmetrically, while preserving their spectral degeneracies\cite{rontgen2021latent}. 
         
         Beyond this, recent studies have uncovered a wealth of physical phenomena in TB models defined on non-Euclidean geometries including topological states\cite{yu2020topological}, quantum computation\cite{kollar2019hyperbolic} and experimental exploration of anti-de Sitter/conformal field theory (AdS/CFT) correspondence\cite{chen2023ads}. The hyperbolic lattice serves as a representative example, with its negative curvature offering a distinct geometric structure compared to Euclidean lattices. Moreover, hyperbolic lattices, constructed via recursive addition of regular polygons to outer edges in negatively curved space and expressed by the Schläfli symbol, support an infinite number of sites\cite{boettcher2022crystallography}. In fact, most hyperbolic TB models are analyzed using graph-theoretic approaches (assigning vertices to elements and graph edges to interelemental interactions) and finite-size exact diagonalization, as illustrated in Fig.~2k\cite{kollar2019hyperbolic}. Therefore, such hyperbolic lattices allow modulation of coupling strength, on-site terms, and effective fields, thereby enabling the realization of unconventional topological phenomena, such as the hyperbolic counterpart of the quantum spin Hall effect\cite{yu2020topological}, higher-order topological states in hyperbolic lattices\cite{zhang2022observation} and hyperbolic Chern insulators\cite{liu2022chern}. Beyond topological physics, hyperbolic lattices also enable on-chip quantum simulation of exotic materials with novel spectral properties, such as flat band comprising a macroscopic number of degenerate states\cite{kollar2019hyperbolic}. Furthermore, recent work has demonstrated the first experimental exploration of AdS/CFT correspondence, by measuring the bulk entanglement entropy (BEE) and boundary-boundary correlation function (BBCF) in hyperbolic lattices, revealing logarithmic scaling consistent with the Ryu–Takayanagi formula and exponential decay governed by the Klebanov-Witten relation, in remarkable agreement with predictions from conformal field theory\cite{chen2023ads}. Collectively, hyperbolic lattices may answer questions at the interface of quantum mechanics, gravity, and condensed matter physics, particularly given that this type of lattices cannot be realized in naturally occurring materials.

		\subsection{Three-dimensional tight-binding models}
		So far, TB models in one and two dimensions have revealed diverse physical phenomena, including quantum Hall effect, quantum spin Hall effect, HOTIs and various types of topological defects. As a meaningful extension, 3D TB models have proven essential for the exploration of novel topological phases, such as 3D topological insulators and topological semimetals. Here, 3D TB models in general could be classified as gapped and gapless. The gapped 3D TB models can be regarded as extensions of lower-dimensional topological structures, such as 3D Chern insulators\cite{stormer1986quantization,liu2022topological}, although their topological characteristics differ from those in lower dimensions. The gapless 3D TB models relies entirely on the topology of 3D momentum-space, and in principle has no lower-dimensional counterpart, such as Weyl semimetals\cite{ghorashi2020higher}.	
		
		As in two dimensions, 3D gapped topological phases can be realized in both TR symmetric and TR symmetry-broken TB models. In TR symmetry-broken cases, a representative example is 3D quantum Hall phase, which is characterized by a triplet Chern vector [$C=(C_x,C_y,C_z)$]\cite{vanderbilt2018berry}. The multicomponent nature of the Chern vector hence generalizes 1D chiral edge states to 2D surface states. Notably, the 3D quantum Hall phase has been experimentally realized in semiconductor superlattices formed by stacking 2D quantum Hall insulators with appropriate interlayer coupling\cite{stormer1986quantization}. In comparison, the 3D Chern PCs with chiral surface states was only recently observed\cite{liu2022topological}. The experimental realization of 3D Chern insulators offer a promising route towards future high-efficiency and low-loss electromagnetic transmission devices. In TR symmetric cases, in a natural parallel with 2D quantum spin Hall effect, the 3D quantum spin Hall effect is characterized by four $Z_2$ invariants $v_0;(v_1 v_2 v_3$)\cite{fu2007topological}. The $v_0$ index classifies 3D quantum spin Hall insulators into weak topological insulator ($v_0=0$) and strong topological insulator ($v_0=1$). For example, a 3D weak topological insulator can be realized by stacking 2D quantum spin Hall insulators similar to 3D Chern insulator, and supports topological surface states on some surfaces\cite{hasan2010colloquium}. Moreover, such 3D topological insulators have also been reported to exhibit large spin Hall angles, and they hold great potential for ultralow-power magnetoresistive random-access memory applications\cite{khang2018conductive}.  
		
		Another crucial 3D gapped TB model is 3D BBH model, whose topological invariants are quantized octupole moments. 3D $n^{th}$-order topological insulator exhibit more protected states at codimension-$n$ (codim-$n$) boundaries: surfaces (codim-1), hinges (codim-2), and corners (codim-3)\cite{benalcazar2017quantized}. As noted in previous subsections, an alternative pathway to realizing 2D HOTIs circumvents the need for negative couplings in BBH model by employing the SSH model. Notably, this approach is also applicable to 3D systems via dimensional extensions. It is well established that the 2D SSH lattice arises from coupling 1D SSH chains via in-plane coupling, while the 3D counterpart is constructed by stacking 2D SSH models with controlled interlayer couplings\cite{liu2023analytic}. For the 3D SSH model, topological properties are characterized by 3D generalized dipole polarizations. Furthermore, diverse 3D lattices exhibit higher-order topology through distinct geometric structures. For instance, the breathing pyrochlore lattice (Fig.~2l) realizes a 3D HOTI consists of tetrahedral units with alternating bond expansions\cite{ezawa2018higher}. Through precisely modulating coupling terms, the breathing pyrochlore lattice is a third-order topological insulator which has four topological boundary states emerge at the corners of the tetrahedron with a $1/4$ fractional charge at each corner. Remarkably, HOTIs can be realized in hetero-structures, such as the twisted bilayer grephenes with large angles host topological corner charges\cite{park2019higher}. Notably, the combination of van der Waals materials with 3D higher-order topological phases provides a platform for future spintronics applications\cite{noguchi2021evidence}. 
		
		On the other hand, 3D gapless TB models exhibit fundamentally different topological phases. These phases, such as Weyl~\cite{ghorashi2020higher}, Dirac~\cite{qiu2021higher}, and nodal line semimetals~\cite{park2022nodal}, have no 2D counterparts. At the heart of topological semimetals lie momentum-space band degeneracies – crossings between two or more bands (Weyl points for twofold degeneracies, Dirac points for fourfold degeneracies) that relate to their nontrivial topological phases. For example, hinge states can be realized in Weyl semimetals described by 3D TB models \cite{xia2022experimental,wei2021higher}. Such Weyl semimetals can be provided by the breathing kagome lattice with helical interlayer coupling terms along $z$-direction\cite{wei2021higher}. Under specific coupling conditions, the twofold band degeneracies emerge at certain momentum points, while their TR counterparts appear at distinct, inequivalent positions in momentum-space. Consequently, Fermi-arc states on 2D surfaces and 1D hinge states emerge in these systems. Weyl points can be reviewed as transition points between trivial insulators and HOTIs parameterized by the wave vector $k_z$ along the $z$-direction\cite{wei2021higher}. However, most experimental realizations remain predominantly limited to acoustic platforms, as photonic systems face inherent challenges in implementing the required helical coupling\cite{xiang2024demonstration,xia2022experimental,wei2021higher}. Beyond higher-order topology, other physical properties of Weyl points also find intriguing applications, such as the chiral bulk modes of Weyl points have been utilized to achieve robust photon transport within the bulk medium\cite{jia2019observation}.

		\subsection{How to achieve tight-binding models in photonics: similarity and difference?}
		The realization of TB models in photonic systems hinges on establishing precise mappings between parameters of the Hamiltonian and tunable photonic aspects. In 2D dielectric PCs, a dielectric rod is commonly used as the structural unit corresponding to a TB lattice. The complex amplitude of strong polarized resonance mode supported by the rod is regarded as scalar field localized at the site of the TB model. This formulation establishes a direct mapping to the TB models: the on-site term in the TB model is mapped by the frequency of the resonance mode, while the hopping term is mapped by the overlap (coupling strength) of resonance modes in adjacent rods of PCs. Hence, altering the on-site term in the TB model can be realized by changing the relative permittivity or structure parameters of the dielectric rods, while modifying the coupling terms correspond to adjusting the distances between the adjacent rods \cite{joannopoulos2008molding,xie2018second,chen2019direct}. This correspondence appears quite seamless, suggesting that many TB models can potentially be realized in photonic systems, facilitating more intuitive manipulation of the internal aspects, include on-site term engineering, coupling term engineering, lattice geometric engineering, effective field engineering and interface engineering. However, the mapping from the resonance modes of dielectric rods to the TB model is not mathematically exact. The field profiles of these resonance modes in PCs decay slowly away from each rod, resulting in long-range coupling between rods that are not nearest-neighbors\cite{li2024disentangled}. On the other hand, the resonance modes treated as scalar fields must exhibit strong polarization along a specific direction, imposing stringent design requirements on the dielectric rods. In the mapping procedure, non-nearest-neighbor couplings and nonideal polarization effects are neglected, and these approximations make discrepancies of the band structure between PCs and TB models. Such discrepancies become even more pronounced in 3D systems, posing a significant limitation on the design of PCs that faithfully realize TB models. Fortunately, advances in technologies such as laser-written waveguides array, coupled resonator optical waveguides array, and artificial hetero-structured PCs may provide promising opportunities for realizing a variety of exotic photonic TB models.
        
        \newpage
		\section{Photonic waveguide arrays}
	    In this section, we focus on the implementation of TB models in photonic waveguide arrays. Photonic waveguides confine light within high-refractive-index media through the principle of total internal reflection. When only a single mode or a specific polarization mode is considered, the light in the waveguides is primarily described by the complex amplitude of the mode, allowing each individual waveguide to be regarded as a vertex in the TB models. The coupling between adjacent waveguides, established via evanescent waves, determines the interaction between vertices. In photonic waveguide arrays, the coupling between multiple waveguides forms a lattice-like system, leading to the emergence of photonic band structure.\cite{longhi2009quantum} Therefore, by engineering diverse waveguide arrays, it is possible to realize various topological phases predicted in TB models. This section first introduces the principles of photonic waveguide arrays, and then reviews recent progress in both 1D and 2D configurations.
	    
		\subsection{Principle of photonic waveguides arrays}
        In 2004, T. Pertsch et al.\cite{pertsch2004discrete} has first demonstrated the fabrication of photonic waveguide arrays using the femtosecond laser direct writing technique. This technology provides a versatile method for realizing TB model, including 1D SSH model\cite{wang2019direct}, honeycomb lattice model\cite{rechtsman2013topological} and square lattice model\cite{cerjan2020observation}. Furthermore, photonic waveguide arrays can also be fabricated using continuous-wave laser writing\cite{xia2018unconventional} and electron-beam lithography techniques\cite{zhang2019experimental}, and the explorations of topological states in non-Hermitian lattice systems are also enabled by using materials with gain and loss\cite{song2019breakup}.
        
       The principle of photonic waveguide arrays can be clearly illustrated by drawing an analogy between the Schrödinger equation and the paraxial Helmholtz equation. In a quantum mechanical setting, the dynamics of a quantum wavefunction $\Psi(x, y, t)$ is typically governed by the Schrödinger equation:
		\begin{equation}
			i\hbar \frac{\partial \Psi(x, y, t)}{\partial t} = - \left( \frac{\hbar^2}{2m} \left[ \frac{\partial^2}{\partial x^2} + \frac{\partial^2}{\partial y^2} \right] - V(x, y, t) \right) \Psi(x, y, t)
		\end{equation}
		In optical systems, the electric field $E(x,y,z)$ of light propagating along the $z$-direction can be described by the paraxial Helmholtz equation:
		\begin{equation}
			i\bar{\lambda} \frac{\partial E(x, y, z)}{\partial z} = - \left( \frac{\bar{\lambda}^2}{2n_0} \left[ \frac{\partial^2}{\partial x^2} + \frac{\partial^2}{\partial y^2} \right] + \Delta n(x, y, z) \right) E(x, y, z)
		\end{equation}
	    The analogy of the two equations becomes evident upon recognizing the correspondence between the key physical parameters: the propagation coordinate $z$ and the evolution time $t$, the reduced wavelength $\bar{\lambda}=\lambda/2\pi$ and the Planck’s constant $\hbar$, the mass $m$ and the refractive index $n_0$, and the refractive index $\Delta{n}$ and the potential $V$. Moreover, the transverse $(x,y)$ plane takes the role of an artificial 1D or 2D material, such as some lattice systems\cite{szameit2010discrete}. Here, each waveguide acts as a meta-atom that represents a site of the TB model, while the evanescent coupling between neighbor waveguides corresponds to the hopping between adjacent sites, as shown in Fig.~3a. As a result, it is possible to employ the TB approximation to describe the light evolution in photonic waveguide arrays, we can write the paraxial equation as\cite{ozawa2019topological}
		\begin{equation}
			i\frac{\partial\psi_m}{\partial z} + \sum\limits_{<m,n>} J_{m,n} \psi_n=0
		\end{equation}
		where $\psi_m$ is the amplitude of the mode in waveguide $m$, $J_{m,n}$ represents the coupling strength and quantifies the evanescent overlap between waveguides $m$ and $n$. Crucially, the coupling coefficient $J_{m,n}$ can be tuned by adjusting the spacing between waveguides, waveguide refractive index and wavelength. Furthermore, the presence of the propagation direction $z$ allows the direct experimental observation of electromagnetic wave dynamics\cite{crespi2013dynamic}. Consequently, a wide range of physical phenomena described by TB models can be simulated by waveguide arrays, such as topological one-way edge states\cite{rechtsman2013photonic}. 
		
		\subsection{One-dimensional photonic waveguide arrays}
		In this subsection, we briefly discuss the realization of 1D TB models in photonic waveguide arrays. The experimental demonstration of the Shockley-like surface states exhibits the first exploration of topological states in the 1D SSH model within photonic systems\cite{malkova2009observation}. Following this pioneering work, the topological phases of 1D TB model have been widely discussed and realized in photonic waveguide arrays. 
		
		The SSH waveguide array, as a representative realization of 1D photonic lattices, exhibits a topological phase transition that can be induced by tuning the spacing between adjacent waveguides. When two SSH-type silicon waveguide arrays with distinct Zak phases are interfaced , a topologically protected zero-energy mode emerges at their boundary (see Fig.~3b)\cite{blanco2016topological}. Furthermore, plasmonic waveguide arrays have been proposed as an alternative platform for observing topological states of 1D SSH model\cite{bleckmann2017spectral}. It is widely believed that any perturbation breaking the topological phase simultaneously destroys the associated topological edge states\cite{chiu2016classification}. However, recent studies have demonstrated the existence of topological edge states protected by chiral symmetry in 1D SSH photonic waveguide arrays, even in cases where the overall topological invariant and the associated topological phase is broken through introducing long-range coupling\cite{wang2023sub}(Fig. 3c). Notably, the existence of topological states in SSH waveguide arrays opens up novel avenues for light manipulation. For example, A. Blanco-Redondo et al.\cite{blanco2018topological} have exploited these states to generate biphoton states and realize topological entanglement.     
		
		Beyond the SSH model, photonic waveguide arrays offer a powerful platform for realizing a broader class of TB models, including those with time-dependent modulations. A pioneering example is the off-diagonal Harper model, which describes a 1D lattice with periodically modulated inter-site coupling\cite{kraus2012topological}. In the photonic implementation, the coupling strength is controlled by the waveguide spacing, and a variation of this spacing along the propagation direction $z$ effectively maps time variations onto spatial variations. As a result, two single-photon topological states, initially localized at the opposite edges of the waveguide array, are adiabatically delocalized to the bulk, where they interfere and undergo a beam splitter operation\cite{tambasco2018quantum}. Crucially, the quantum interference of single photons is a foundational mechanism in linear optical quantum computation\cite{o2009photonic}. Another intriguing TB model is the Aharonov-Bohm (AB) cage chain, a quasi-1D lattice composed of interconnected plaquettes\cite{mukherjee2018experimental}. As a pioneering implementation, M. Kremer et al.\cite{kremer2020square} realized this model in photonic waveguide arrays by introducing an auxiliary waveguide between two original waveguides to generate an effective negative coupling. This tailored structure enables the realization of a square-root topological insulator, which hosts robust boundary states characterized by a non-quantized topological index. Furthermore, 1D models with multiple topologically nontrivial bands can be implemented in photonic waveguide arrays. The coexistence of multiple states at the same edge gives rise to intriguing edge dynamics, which can be validated through a characteristic spatial beating effect\cite{zhang2019experimental}.
		
		Moreover, a variety of physical phenomena associated with 1D non-Hermitian TB model have been widely studied in the photonic waveguide arrays. For example, the breakup of zero modes in finite SSH model can be recovered by introducing gain and loss into the silicon waveguides\cite{song2019breakup}, as illustrated in Fig. 3d. Since common materials used for photonic waveguides (e.g., silicon and silica) cannot directly provide optical gain, related research has primarily focused on passive non-Hermitian systems, where the extra loss is introduced by patterning dissipative metal such as chromium (Cr)\cite{feng2013experimental,song2019breakup} or spatially periodically “wiggling” waveguides\cite{weimann2017topologically}. By carefully designing loss in the silicon waveguide arrays, both the parity-time (PT) and broken-PT phases can be realized. Notably, although the zero mode usually deviates from the exact zero energy state in a finite Hermitian lattice due to the coupling between these edge states, they can be restored by introducing non-Hermitian degeneracies with broken-PT symmetry. Similarly, recent studies have demonstrated that the precise manipulation of on-site gain and loss in silicon waveguide arrays enables active control over both the topological phase transitions and the PT phase transition of the photonic systems\cite{pan2018photonic}. Furthermore, optical nonlinearities play a pivotal role in 1D TB model with PT symmetry. When the nonlinearity exceeds a certain threshold, the system can undergo a transition from the PT-broken phase to the PT-symmetric phase\cite{lumer2013nonlinearly}. For instance, S. Q. Xia et al.\cite{xia2021nonlinear} demonstrated self-focusing or self-defocusing nonlinearities can be used to tune the effective loss in waveguides. Hence, nonlinear excitation of the interface waveguide can affect the properties of the whole lattice, leading to the transition between PT-symmetric and non-PT-symmetric regimes as well as the control of topological phase. Importantly, the ability to manipulate global topological properties and PT symmetry through optical nonlinearity offers promising opportunities for unconventional control of light propagation\cite{xia2021nonlinear}. 
		
		As discussed in the previous section, the NHSE represents a hallmark phenomenon in non-Hermitian physics. However, realizing the NHSE in photonic waveguide arrays poses a fundamental challenge due to the inherent difficulty in establishing the required non-reciprocal couplings\cite{weidemann2020topological}. Furthermore, the NHSE can also be observed in Floquet-engineered non-Hermitian photonic waveguide arrays without non-reciprocal coupling. For instance, Floquet NHSE of light can be achieved by engineering the interplay between synthetic gauge fields and optical loss in coupled waveguides fabricated on a silicon platform. Moreover, a topological phase transition between unipolar and bipolar regimes can be induced by tuning the modulation periodicity and the dissipation strength\cite{lin2024observation}. Nevertheless, it is not impossible to achieve non-reciprocal coupling in the photonic waveguide arrays. Remarkably, it has been shown that a phase lag induced by loss, which is independent of the energy propagation direction, results in different interference outcomes for the forward and backward directions\cite{huang2021loss}. Building upon this mechanism, non-reciprocal coupling can be designed by introducing two auxiliary waveguides with on-site dissipation, thereby establishing two distinct coupling pathways between adjacent primary waveguides and an artificial gauge field. As a result, the NHSE can emerge from the interplay between the synthetic gauge field and dissipation\cite{ke2023floquet}. Notably, the phase transition mechanism between NHSE and topological states enables the realization of a topological switch for the skin-topological effect\cite{sun2024photonic}.

		\begin{figure}[H]
			\         	\setlength{\abovecaptionskip}{-15pt}
			\begin{center}
				\begin{tabular}{c}
					\includegraphics[height=15cm]{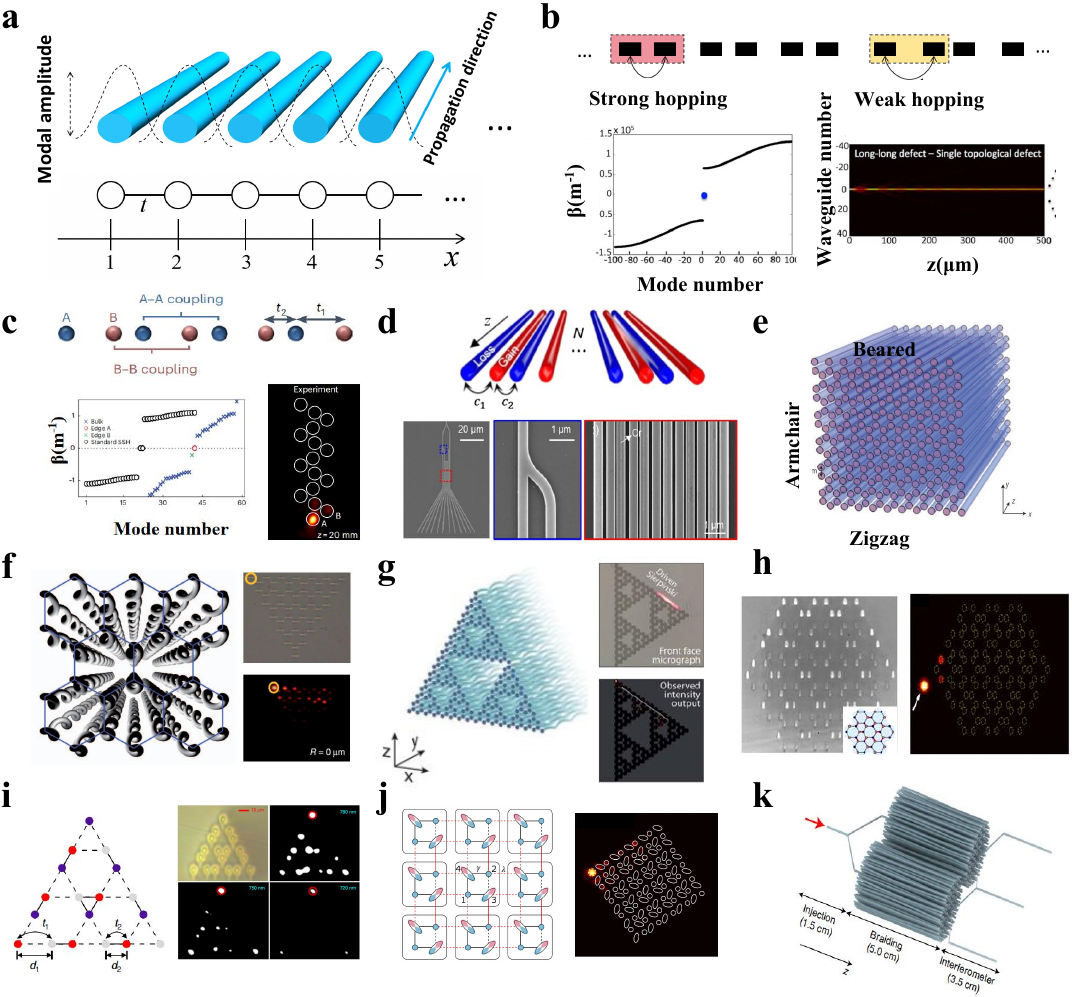}
				\end{tabular}
			\end{center}
			\caption 
			{ Realizations of various TB models in 1D and 2D photonic waveguide arrays. \textbf{a}. Schematics of 1D waveguides array. The transverse ($x$) plane takes the role of an artificial material and $z$-direction is propagation direction. Dashed lines represent modal amplitude distribution in a waveguide. \textbf{b}. The top panel shows the connection of two SSH models with distinct phases. The bottom panel shows the eigenspectra of the finite waveguide array (left) and light propagation in the structure when zero mode is excited (right). Figures adapted from Ref.~\cite{blanco2016topological} . \textbf{c}. The top panel illustrates the eigenspectra of the 1D SSH model with and without next-nearest-neighbor coupling. The bottom panel shows the experimental realization using a photonic waveguide array, demonstrating the sub-symmetry protection of edge modes. Figures reprinted from Ref.~\cite{wang2023sub} . \textbf{d}. Schematics of the 1D non-Hermitian SSH model with $N$ waveguides, where red and blue rods represent the gain and loss respectively (top), and SEM image of the fabricated structure (bottom). Figures reprinted from Ref.~\cite{song2019breakup} . \textbf{e}. Schematics of honeycomb lattice in photonic waveguides array, $z$-direction is the propagation direction. Figure adapted from Ref.~\cite{plotnik2014observation} . \textbf{f}. Geometry (left) and experimental results (right) of honeycomb photonic Floquet topological helical waveguides array, where the yellow circles shows the position of injection light. Figures reprinted from Ref.~\cite{rechtsman2013photonic} . \textbf{g}. Schematics of Penrose tiling waveguides array (left) and its topological edge transport (right). Figures reprinted from Ref.~\cite{bandres2016topological} . \textbf{h}-\textbf{j}. Experimental realizations of honeycomb lattice (\textbf{h}), kagome lattice  (\textbf{i}) and BBH model (\textbf{j}), along with the corresponding observations of topological corner states shown in the right panel of each diagram. Figures adapted from Ref.~\cite{noh2018topological} , Ref.~\cite{el2019corner} and Ref.~\cite{schulz2022photonic} . \textbf{k} Schematics of the waveguides array implementing non-abelian braiding.    Figure reprinted from Ref.~\cite{noh2020braiding} . 
				}
		\end{figure} 

		\subsection{Two-dimensional photonic waveguide arrays}
		Since the proposal of the photonic quantum Hall effect by Haldane and Raghu\cite{haldane2008possible}, this idea has been rapidly realized in practice through creating a gyromagnetic PC in the microwave range\cite{wang2009observation}. However, magnetic effects at optical frequencies are notoriously weak, making it difficult to break TR symmetry through magneto-optical devices. Since photonic waveguides map temporal evolution onto spatial variations, breaking TR symmetry can be achieved by introducing spatial inversion asymmetry  along the propagation direction of light. Over the past decade, photonic waveguide arrays, as a significant class of photonic systems, play a pivotal role  in the realization of 2D TB models.
		
		The experiment realization of 2D TB lattice in photonic waveguide arrays was proposed by M. C. Rechtsman et al.\cite{rechtsman2013strain}, who realized a 2D honeycomb photonic lattice. In a follow-up work, they further investigated the edge states supported by zigzag and bearded boundaries of the honeycomb photonic lattice, as shown in Fig.~3e\cite{plotnik2014observation}. Crucially, the propagation coordinate $z$ in this system acts as a temporal coordinate, such that breaking $z$-reversal symmetry is equivalent to break TR symmetry. Therefore, M. C. Rechtsman et al.\cite{rechtsman2013photonic} successfully realized topologically protected chiral edge states through the implementation of helical waveguide arrays, as shown in Fig.~3f. Moreover, the topological properties of the photonic waveguide arrays are still characterized by a nonzero Chern number, for which these systems with periodic driving are also referred to as Chern-type Floquet topological insulators. Intriguingly, photonic Floquet topological insulators can robustly protect the transport of quantum information through photonic networks\cite{rechtsman2016topological}.
		
		Subsequently, two independent experimental studies have demonstrated a new category of Floquet topological insulators in photonic waveguide arrays, known as anomalous Floquet topological insulators\cite{maczewsky2017observation}. This structure using helical waveguide arrays designed according to the 2D "Rudner-toy" model\cite{rudner2013anomalous}, in which the topological phase transitions can be tuned by varying the waveguide spacing along the propagation axis. Remarkably, despite possessing a vanishing Chern number, these anomalous Floquet topological insulators support robust chiral edge states. Furthermore, similar helical waveguide arrays can be utilized to realize a topological Anderson insulator by introducing disorder. Specifically, the introduction of on-site disorder, implemented as random variations in the refractive index of individual waveguides, can drive the systems from a trivial phase into a topological one. The experimental result hence exhibits the disorder can enhance transport rather than arrest it\cite{stutzer2018photonic}. Notably, the $z$-modulated photonic waveguide arrays can also realize a 2D topological pump through constructing an inter-waveguide separation. The pump can be used to probe the edge states of a 4D quantum Hall system, which incorporates two additional synthetic dimensions\cite{zilberberg2018photonic}. It is worth mentioning that fractal lattice (Fig. 3g)\cite{bandres2016topological, biesenthal2022fractal} with helical waveguide arrays have been proposed. Remarkably, these aperiodic systems can support a variety of topologically protected chiral edge states, even in the absence of bulk bands. Additionally, light transport in fractal system features increased velocities compared with the corresponding honeycomb lattice\cite{biesenthal2022fractal}. 

        In addition to Floquet insulators, photonic waveguide arrays also provide an ideal platform for realizing topological states of 2D TR symmetric TB models. A typical example is: Rechtsman et al.\cite{noh2018observation} experimentally demonstrate the topological valley-Hall edge states using straight waveguide arrays. By detuning the refractive indices of each waveguide, inversion symmetry is explicitly broken, giving rise to zigzag edge states localized along domain walls' separating regions characterized by opposite valley Chern numbers\cite{zhang2011spontaneous}. Furthermore, photonic waveguide arrays have also enabled the experimental realization of HOTIs. For example, J. Noh et al.\cite{noh2018topological} first demonstrated mid-gap 0D corner states protected by chiral symmetry in a $C_{6v}$-symmetric photonic hexagonal waveguide arrays (Fig.~3h), with the topological phase controlled by tuning the inter-waveguide spacing. Similar 0D corner states have subsequently been observed in various other TB models implemented via laser-written waveguide arrays, including the breathing Kagome lattice (Fig.~3i) \cite{el2019corner}, square lattice geometries \cite{cerjan2020observation, wang2021quantum}, and fractal lattices\cite{ren2023theory}. Among HOTI models, the BBH model is particularly challenging to realize in photonic systems due to its requirement of effective negative coupling. To address this challenge, J. Schulz et al.\cite{schulz2022photonic} recently proposed that the mixing of orbitals with distinct parity representations is useful for generating systems with alternating phase patterns. As a result, the BBH model can be realized by leveraging both $s$ and $p$ orbital-type modes, as illustrated in Fig.~3j. Moreover, topological corner states associated with $p$-orbital modes have been experimentally observed in photonic Kagome waveguide arrays\cite{zhang2023realization}. Remarkably, recent study has highlighted the interplay between vorticity, disclinations and higher-order topological features in photonic waveguide arrays, enabling robust transport of optical vortices\cite{hu2025topological}. 
        
        Another compelling direction lies in the realization of non-Abelian phenomena in photonic waveguide arrays. At the core of these phenomena is the emergence of non-Abelian geometric phases, characterized by matrix-valued phases acquired through adiabatic and cyclic evolution in parameter space\cite{bohm2013geometric}. These phases reflect the internal degrees of freedom and path-dependent evolution of the system’s wavefunction, distinguishing them fundamentally from Abelian Berry phases. In particular, non-Abelian braiding operations have attracted considerable attention owing to their prospective applications in topological quantum computation\cite{nayak2008non}. A recent study\cite{noh2020braiding} has proposed the realization of braiding in the dynamic evolution of two zero vortex modes, implemented in photonic honeycomb waveguide arrays under Kekulé modulation\cite{menssen2020photonic}, as shown in Fig.~3k. However, a key challenge lies in the fact that non-Abelian braiding requires the operation of at least three degenerate states. Realizing and controlling multiple topological vortex modes in such large-scale photonic waveguide arrays thus presents substantial experimental difficulties, particularly due to fabrication constraints. Nevertheless, non-Abelian braiding can be realized by enabling multiple single-photon modes in meandering photonic waveguide arrays to undergo adiabatic evolution, thereby inducing a purely geometric-phase effect via tuning the inter-waveguide separations\cite{zhang2022non}. Furthermore, Y. K. Sun et al. have demonstrated non-Abelian Thouless pumping in Lieb lattice consisting of photonic waveguides, in which a flat band exists at zero energy. By modulating the coupling coefficients within waveguides supporting degenerate flat-band modes, they drive the spatial evolution of the guiding modes. Notably, non-Abelian Thouless pumping was experimentally observed in a three-step pumping device, where the final outcomes explicitly depend on the sequence of the pumping operations\cite{sun2022non}. Collectively, non-Abelian operations of photons provide new toolsets for realizing topological quantum logic gates on bosonic platforms\cite{yang2024non}.

		\newpage
		\section{Coupled resonator optical waveguide arrays}
		In this section, we review the realization of several TB models in the coupled resonator optical waveguide (CROW) arrays, including both 1D and 2D systems. Notably, the CROW system have exhibited potential in a wide variety of applications, ranging from optical filtering to slow light delay lines, from nonlinear signal processing to trapping of light in the past decades\cite{morichetti2012first}. With the rapid advancement of topological photonics, the CROW systems endowed with topological phases have enabled various superior applications, such as quantum light generation, optical isolation and topological laser\cite{bandres2018topological}. 
		\subsection{Principle of coupled resonant optical waveguide array}
		In general, CROW arrays are composed of a series of ring resonators. The ring resonator can be regarded as a closed-loop optical waveguide, in which only those wavelengths that satisfy the resonant interference condition are selected and sustained within the cavity\cite{bogaerts2012silicon}. To illustrate the principles of CROW arrays, an infinite 1D array of coupled ring resonators is considered, as depicted in Fig.~4a. Each resonator supports both clockwise (CW) mode and counterclockwise (CCW) mode. Here, it is assumed that the CW and CCW modes within each resonator do not interact, and that the CW mode of one ring couples only to the CCW mode of the adjacent ring. Using the notation of Fig.~4a, the amplitude relationship between two adjacent resonators can be described by\cite{yariv2000universal} 
			\begin{equation}	
\begin{bmatrix}
 a_{n+1} \\ d_n
\end{bmatrix}
=
\begin{bmatrix}
 \eta &  \kappa\\
 -\kappa^* & \eta ^*
\end{bmatrix}
\begin{bmatrix}
 b_{n+1} \\ c_n
\end{bmatrix}
			\end{equation}	
		where $\eta $ and $\kappa$ are the dimensionless transmission and coupling coefficients, respectively. The matrix is unitary and unimodular, i.e.,  $|\eta |^2 + |\kappa|^2 = 1$, so that we obtain:
				\begin{equation}
\begin{bmatrix}
 a_{n+1} \\ b_{n+1}
\end{bmatrix}
=
 \frac{1}{\kappa^*}\begin{bmatrix}
 1 &  -\eta \\
 \eta ^* & -1
\end{bmatrix}
\begin{bmatrix}
 c_n \\ d_n
\end{bmatrix}
=M
\begin{bmatrix}
 c_n \\ d_n
\end{bmatrix}
		\end{equation}
		However, as the optical field circulates within ring, it accumulates a phase shift and may be attenuated. To account for this effect, the field amplitude $[c_n,d_n]^T$ can be described by
		\begin{equation}
\begin{bmatrix}
 c_{n} \\ d_{n}
\end{bmatrix}
=
 \begin{bmatrix}
 0 &  e^{-i\beta R \pi}\\
 e^{i\beta R \pi} & 0
\end{bmatrix}
\begin{bmatrix}
 a_n \\ b_n
\end{bmatrix}
=Q
\begin{bmatrix}
 a_n \\ b_n
\end{bmatrix}
		\end{equation}
		Here, $R$ is the ring radius, $\beta = n_0\omega/c + i\gamma$, with $n_0$ denoting the refractive index, $\omega$ denoting the frequency and $\gamma$ representing the loss (or gain) per unit length in the ring\cite{poon2004matrix}. Therefore, the amplitudes can be connected by the transfer matrix. By subsequently applying the Bloch boundary condition, the eigenvalue equation governing the band structure of the lattice can be expressed as
				\begin{equation}
		MQ\begin{bmatrix}
 a_{n} \\ b_{n}
\end{bmatrix}
=
 e^{ik}
\begin{bmatrix}
 a_n \\ b_n
\end{bmatrix}
		\end{equation}
		
        Therefore, in the CROW arrays, light propagation is governed by effective TB Hamiltonians, which describe the evanescent coupling of light between adjacent resonators\cite{yariv1999coupled}. By adjusting the separation between resonant rings, the coupling coefficient can be effectively modulated, while introducing antiresonant links between rings allows for tuning the phase of coupling. Collectively, these ingredients make the CROW a suitable platform for exploration and realization of topological phenomena based on TB models\cite{liang2013optical}. It is noteworthy that the analysis of CROW involves approaches such as the TB approximation\cite{yariv1999coupled} and temporal coupled mode theory\cite{reynolds2001coupled}. Depending on the specific problem, it is crucial to select the most suitable method.
		
		\subsection{One-dimensional coupled resonant optical waveguide array}
		The first experimental realization of a ring resonator was reported by R. G. Walker and C. D. W. Wilkinson in 1983, using a silver-ion-exchange glass-based waveguide\cite{walker1983integrated}. However, the concept of a CROW was not introduced until 1999, when A. Yariv et al.\cite{yariv1999coupled} formally proposed the term and theory. Approximately a year later, the first experimental demonstration of a CROW structure was achieved using coupled defects in a PC waveguide\cite{olivier2001miniband}. Since then, CROWs have been realized using various cavities including ring resonators\cite{xia2007ultracompact}, square resonators\cite{hafezi2013imaging} and PC cavities\cite{han2019lasing}. In particular, topological designs have significantly enhanced the robustness and performance of ring resonator-based devices\cite{mittal2018topological}.
		
		We begin by introducing the realization of 1D TB models and their associated topological phenomena in CROW arrays. In 2018, M. Parto et al.\cite{parto2018edge} experimentally implemented the SSH model using a 1D array of 16 identical microring resonators fabricated on InGaAsP quantum wells. A topologically nontrivial phase was realized by alternating the separation between adjacent rings, thereby modulating the coupling strength. By pumping one of the two sublattices, mid-gap topological edge states were successfully observed. Moreover, the robust zero-energy mode can also be realized in on-chip silicon microlaser\cite{zhao2018topological, li2023topological}. In fact, most studies of 1D CROW arrays based on TB models have thus far focused on active photonic systems, which involve energy exchange between the system and the external environment, yielding non-Hermiticity of the system\cite{parto2018edge,zhao2018topological,li2023topological}. This focus is largely driven by potential applications, as the judicious use of non-Hermiticity enables single edge-mode lasing, offering a highly efficient optical source. Such sources are particularly promising for integrated silicon photonics, where robustly feed power is essential for chip-scale communication and computing\cite{jorgensen2022petabit}. Hence, it is worthwhile to explore the fundamental aspects of the interplay between non-Hermiticity and topology in active CROW arrays. 
		
		As an example of silicon microlasers, non-Hermiticity can be realized by generating the gain/loss distribution. Specifically, a layer of approximately 10nm chromium (Cr) is deposited on every second resonator in an array consisting of an odd number of resonators, as illustrated in Fig.~4b\cite{zhao2018topological}. Notably, the topological zero mode can fully exploit the distributed gain domains, since the edge state resides only on the gain site. As a result, it is selectively amplified by the engineered gain distribution, enabling robust single-mode lasing that remains stable even under structural perturbations. Furthermore, the non-reciprocal coupling can be induced by inserting an optical gain and loss medium into the link rings. Here, the CW mode in a site ring only excites the CCW mode in the link ring. Consequently, the clockwise mode of the left site ring couples to the right site ring through the lower semicircle of the link ring. In contrast, coupling in the reverse direction occurs via the upper semicircle. The non-reciprocal coupling is inherently introduced by the presence of gain and loss in the link rings. As a result, by tuning the coupling coefficient, the field distribution becomes localized at one end of the finite 1D CROW array, thereby exhibiting the NHSE, as shown in Fig.~4c\cite{zhu2020photonic}. Moreover, non-reciprocal coupling based on the similar method also holds great promises for achieving on-way transport. For instance, S. Longhi et al.\cite{longhi2015non} proposed the unique unidirectional transport by introducing non-reciprocal gauging in HN chains realized in CROW array. However, experimentally realizing non-reciprocal coupling in CROW arrays remains a significant challenge. It was not until 2022 that the first experimental realization of HN lattices in CROW array was reported, achieved using active optical oscillators that feature both non-Hermiticity and nonlinearity\cite{liu2022complex}. Remarkably, these studies illustrate that CROW arrays incorporating non-Hermitian physics may offer a feasible platform for designing optical devices, such as optical circuits, lasers, and beyond\cite{ota2020active}. 
		
        \begin{figure}[H]
         	\setlength{\abovecaptionskip}{-15pt}
        	\begin{center}
	        \begin{tabular}{c}
		          \includegraphics[height=14.5cm]{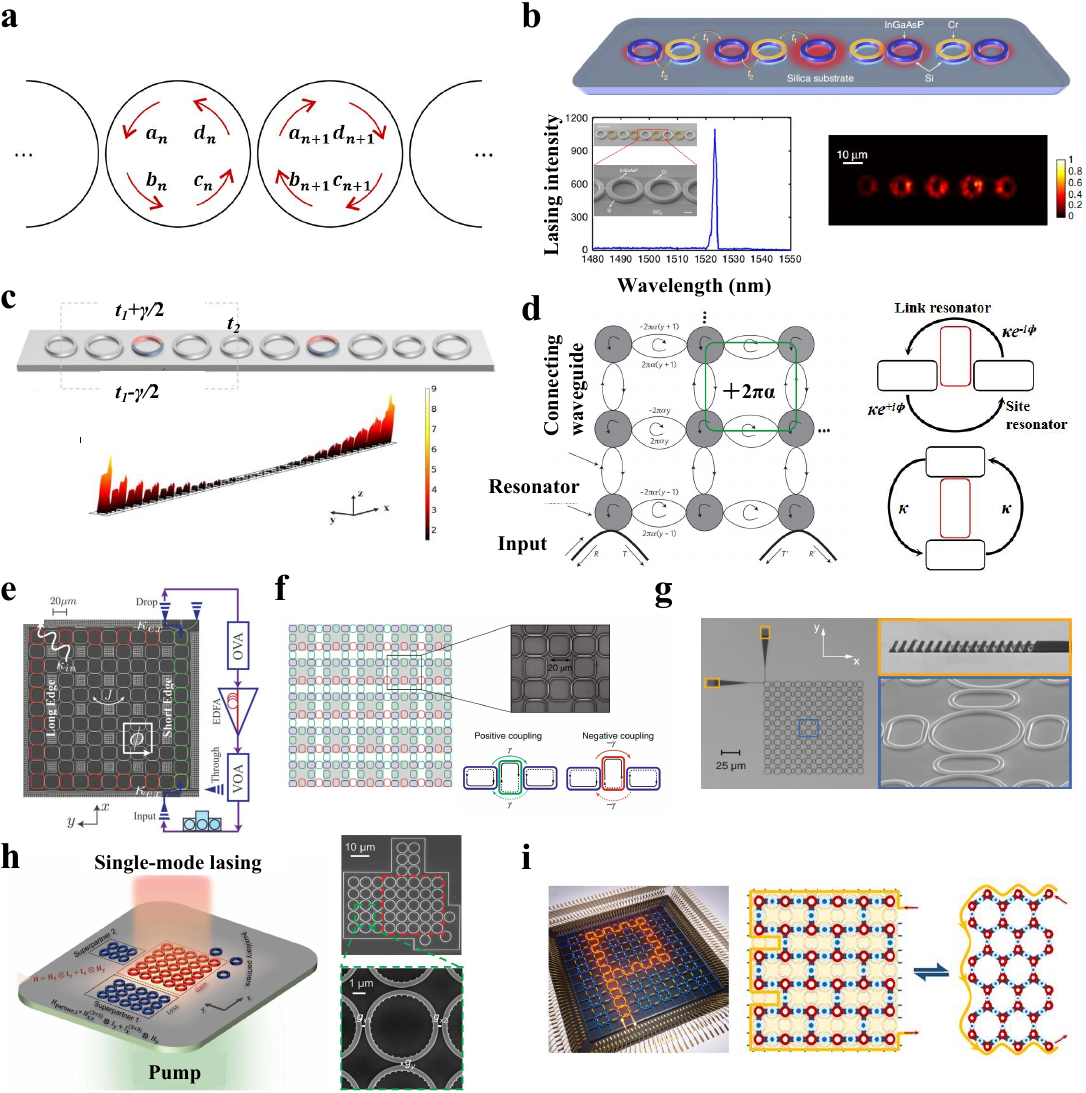}
         	\end{tabular}
            \end{center}
			\caption 
			{Realizations of TB models in 1D and 2D CROW arrays. \textbf{a}. Schematic illustration of coupling between the two coupled ring resonators. $a_{n/n+1}$, $b_{n/n+1}$ and $c_{n/n+1}$, $d_{n/n+1}$ represents the clockwise (CW) and counterclockwise (CCW) propagation field components respectively. Figure reproduced from Ref.~\cite{liang2013optical} . \textbf{b}. Schematic of the CROW array with alternating weak and strong couplings, emulating the SSH model (top). The bottom panel shows the measured lasing emission from the interface state (left), and the spatial profile of the laser emission, which is localized around a topological interface state (right). Figures adapted from Ref.~\cite{zhu2020photonic} . \textbf{c}. Schematic of a CROW array model with non-reciprocal coupling (top), and the realization of the NHSE (bottom). Figures adapted from Ref.~\cite{zhu2020photonic} . \textbf{d}. A 2D array of CROWs realizing the the Harper-Hofstadter model (left). Schematic of the coupling mechanism between two site resonators via a link ring (right). Figures adapted from Ref.~\cite{hafezi2011robust} . \textbf{e}. Experimental setup corresponding to \textbf{d}. Figures reprinted from Ref.~\cite{hafezi2013imaging} . \textbf{f}. Realization of the BBH model in a CROW system, where negative coupling is achieved by tuning the coupling phase via an off-resonant link ring. Zoom-in picture (upper right corner) represents the microscope image of the plaquette. Figures reprinted from Ref.~\cite{mittal2019photonic} . \textbf{g}. The left panel shows an SEM image of a 2D CROW array implementing the Harper-Hofstadter model. Zoom-in pictures of yellow area and blue area indicate the outcoupling grating structures used to probe the array and unit cell consists of a primary ring site surrounded by 4 identical intermediary racetrack links. Figures reprinted from Ref.~\cite{bandres2018topological} .  \textbf{h}. Schematic of a higher-dimensional supersymmmetric microlaser array, consisting of a 5$\times$5 main array (red area), coupled with its two dissipative superpartner arrays and three auxiliary partner microrings (blue area). The left panel shows SEM image of the CROW array. Figures adapted from Ref.~\cite{qiao2021higher} . \textbf{i}. Conceptual diagram of programmable topological photonic circuit devises on a 2D coupled-microring resonators and nanophotonic circuits (left). The arbitrary lattice is realized by integrated Mach-Zehnder interferometer (right). Figures adapted from Ref.~\cite{dai2024programmable} .
			} 
		\end{figure} 
		
		\subsection{Two-dimensional coupled resonant optical waveguide array}
		While the 1D TB model realized in ring resonator systems has enabled a variety of exotic topological phenomena, the first realization of topological photonic ring resonator can be traced back to the original work of M. Hafezi et al.\cite{hafezi2011robust} in 2011, where a 2D ring resonator array was utilized. In this work, TR symmetry breaking was achieved by selectively exciting modes with a specific circulation direction. Subsequently, asymmetric link rings were used to implement a photonic analog of the quantum Hall lattice model in a 2D CROW array. Specifically, the link rings oriented along the $y$ direction were symmetrically positioned to get zero hopping phase. On the other hand, those along the $x$ direction were asymmetrically positioned, in an increase order in the $y$ direction, giving rise to a hopping phase (Fig.~4d). As a result,  this synthetic vector potential is formally equivalent to an out-of-plane effective magnetic field\cite{hafezi2011robust}. Building on this design, the first experimental realization of the lattice model was reported by M. Hafezi et al.\cite{hafezi2013imaging} using standard silicon-on-insulator technology working in the telecom band ($\lambda~=1550$nm). As shown in Fig.~4e, when the CROW system is excited through an input waveguide with a laser field tuned to the frequency of the edge modes, a topological edge state emerges. Notably, the propagation of the edge state is unidirectional and remains robust against disorder. Furthermore, the flexibility of CROW architectures offers a promising platform for topological quantum computation\cite{kavokin2022polariton}, as complex lattices can be constructed by simply interconnecting ring resonators and manipulating the topological states.
		
		Subsequently, S. Mittal et al.\cite{mittal2014topologically} provided a more quantitative experimental demonstration of the structural robustness by employing the same CROW array. Their results confirmed that these edge states support ballistic light transport, exhibiting minimal device-to-device variation in photon delay times. In contrast, conventional CROWs exhibited strong fluctuations due to the disorder induced scattering. In a follow-up, S. Mittal et al.\cite{mittal2016measurement} were the first to experimentally measure the winding number in a 2D CROW lattice, thereby revealing the topological properties of the system. To model the spectral flow of a quantum Hall edge of winding number $k=1$, one considers a linear edge dispersion. When a gauge flux is coupled to the edge, the momentum is replaced by the covariant momentum. For noninteracting photons, the insertion of $2\pi$ flux shifts quantization of momentum $E_n$ to $E_{n-1}$, resulting in a spectral flow. Experimentally, the coupling was realized using integrated metallic heater fabricated above the link rings on the lattice edge. As the coupled flux increase to $2\pi$, the edge state resonances move by one resonance to replace the position once held by its neighbor, giving rise to the measured winding number $k=1$. In addition to the aforementioned works, it has been demonstrated that introducing next-nearest-neighbor couplings in CROW arrays can also lead to quantum Hall-like edge states\cite{leykam2018reconfigurable,mittal2019anomalous}. Notably, the realization of TB models with long-range interactions in photonic systems has been shown to break channel-bandwidth limitations\cite{kim2024long}. Furthermore, a 2D ring resonator array utilized the Haldane model, where a pump localized to the lattice edges induced lasing in its chiral edge states. By adding gain only to the perimeter of the system, lasing preferentially occurs in the topologically protected edge modes. This study is expected to offer an avenue to make many semiconductor lasers operate as one single-mode high-power laser\cite{harari2018topological}. 
		
		In addition to models that rely on TR symmetry breaking, intriguing TB models that preserve TR symmetry have also been realized. For example, the honeycomb lattice can be designed by CROW photonic structures. Nontrivial bandgaps are achieved through utilizing effective gauge fields implemented by the intrinsic pseudo-spin-orbit interaction so that the system hosts robust edge states\cite{zhu2018z}. Another notably example is the realization of quadrupole topological insulators in CROW systems. Leveraging the flexibility of coupling phase engineering in CROWs, it becomes feasible to implement both positive and negative coupling between resonators. Specifically, the strength and sign of the coupling were determined by vertically shifting the link rings between adjacent resonators, as shown in Fig.~4f. Consequently, the coexistence of positive and negative couplings leads to a nonzero quadrupole moment, which in turn gives rise to topological corner states--a hallmark of HOTIs\cite{mittal2019photonic}. Simultaneously, higher-order topological states have also been realized in 2D CROW arrays by introducing staggered on-site gain and loss to a Hermitian trivial system\cite{luo2019higher,ao2020topological}. In this non-Hermitian TB models, the topological invariants are characterized using the biorthogonal nested-Wilson-loop and the edge-polarization formalism. Furthermore, CROW arrays that preserve TR symmetry also hold promise for topological laser applications. For instance, a topological laser operation without need for magnetic elements in a CROW array has been reported, as shown in Fig.~4g. By selectively pumping only the edge resonators, highly efficient single-mode lasing was achieved in the topologically protected edge states, even at gain values high above threshold. Notably, lasing via topological edge states exhibited enhanced propagation efficiency and output power compared to lasing in the perimeter modes of the trivial model\cite{bandres2018topological}. Furthermore, to overcome the scalability challenges in photonic integrated circuits, supersymmetric design principles have recently been introduced into coupled microring lasers. As illustrated in Fig.~4h, a supersymmetric 2D microring laser array was constructed, where all microrings are evanescently coupled to form a well-controlled TB lattice. By carefully engineering the system based on higher-dimensional supersymmetry formalism, all microring elements were controlled precisely, enabling single-frequency laser emission\cite{qiao2021higher}. 
		    
		While drawing significant attention on topological laser, the 2D CROWs offer substantial potential for the implementation of dynamically reconfigurable topological photonic lattices in photonic chip\cite{capmany2020programmable}. A key advantage of programmable CROW arrays lies in their design flexibility, enabling rapid reconfiguration of the photonic circuit to realize a wide range of TB models. Recently, T. X. Dai et al. \cite{dai2024programmable} reported a highly programmable topological photonic chip which integrates large-scale silicon photonic nanocircuits and microresonators, as conceptually illustrated in Fig.~4i. Remarkably, each photonic artificial atom and its interactions can be individually controlled, allowing arbitrary tuning of structural parameters and geometrical configurations. This enables the observation of dynamic topological phase transitions and the realization of various TB models on a single platform. Using this system, authors demonstrated a series of intriguing phenomena, such as real-space distributions of electromagnetic field that are flexibly reprogrammed to display "$PKU$", the coupling-strength-controlled phase transitions\cite{li2009topological}, statistical verification of topological phenomena\cite{hafezi2011robust,mitchell2018amorphous,jia2023disordered}, and realization of photonic insulators with different geometric TB models\cite{slager2013space}. In parallel, another work has proposed reconfigurable topological Hamiltonians by utilizing a reprogrammable integrated photonics platform, consists of a hexagonal mesh of silicon Mach-Zehnder interferometers with phase shifters. Notably, this photonic CROW array showcases its versatility by realizing diverse TB models, such as the 1D SSH model featuring robust edge states and the 2D breathing kagome lattice hosting topological corner states\cite{on2024programmable}. Intriguingly, these advances in CROW array provide a flexible platform for emulating topological TB models and artificial nanomaterials, opening an era of software-defined topological photonics\cite{capmany2024programming}.

		\newpage
		\section{Photonic systems with synthetic dimension}
        As discussed in the previous two sections, both photonic waveguide arrays and CROW arrays utilize the spatial propagation direction of light as an effective extra dimension. Consequently, they typically require the construction of higher-dimensional physical structures to emulate lower-dimensional topological states, result in large-scale system architectures and is incompatible with integrated photonics. A compelling direction to overcome this limitation lies in the photonic systems with synthetic dimensions. The key concept is to reinterpret internal degrees of freedom as spanning additional spatial dimensions, thereby enabling lower-dimensional photonic systems to simulate the topological behaviors found in higher dimensions\cite{ozawa2019topological}. As it turns out, photonic systems with synthetic dimensions have shown significant potential for on-chip device applications\cite{zhu2021integrated}, such as electro-optic microcombs on thin-film lithium niobate\cite{hu2020realization}. In this section, we first discuss the principle of photonic systems with synthetic dimensions, before reviewing realizations of various TB models within these platforms.  		
		
		\subsection{Principle of photonic systems with synthetic dimension}
		The realization of certain 3D TB models in photonic systems remains challenging, primarily due to the structural constraints inherent to photonic waveguides and CROWs. In contrast, photonic systems with synthetic dimensions offer a controllable platform for investigating the physical processes\cite{peruzzo2010quantum}, and have enabled the realization of 3D and higher dimensional TB models \cite{ozawa2019topological}. More generally, by combining synthetic dimensions with the geometric dimensions of a photonic lattice, the effective spatial dimensionality of the system can exceed the physical dimensionality of the real space in which the lattice is located. In photonics, one natural idea for realizing synthetic dimensions is to increase the connectivity of the lattice. For example, the introduction of long-range coupling to a 1D chain with nearest-neighbor interactions can effectively give rise to higher-dimensional TB model\cite{yuan2018review, cheng2023multi}. Another strategy involves exploiting diverse degrees of freedom, including distinct frequency modes\cite{zhang2017generation}, different orbital angular momentum states\cite{luo2017synthetic} and lattice Fock states\cite{zhang2025synthetic}. 
        
        To provide a concrete and rigorous understanding of photonic systems with synthetic dimensions, we illustrate the principle by using synthetic frequency dimension as a representative example. In photonic structures that support multiple discrete frequency modes, it is essential to implement mechanisms that enable coupling between these modes, thereby realizing a TB model along the synthetic frequency dimension. For example, L. Q. Yuan et al.\cite{yuan2016photonic} have proposed to use different frequency modes of a multimode ring resonator, coupled via dynamic modulation of refractive index, as a synthetic dimension. As an illustration, consider a static ring resonator equipped with a time-dependent refractive index modulator, such as an electro-optic modulator (EOM). Here, we assume the single-mode waveguide forming the ring exhibits zero group velocity. This ensures that resonant modes form an equally spaced frequency comb, where the number of modes within this near-zero group velocity dispersion range determines the synthetic lattice size along the frequency dimension. If the resonator supports a mode at frequency $\omega_0$, the frequency of the $m$th resonant mode is given by $\omega_m = \omega_0 + m\Omega_R$, where the resonant modes form an equally spaced frequency comb. Here, $\Omega_R = 2\pi c / n_0 L$ denotes the free spectral range (FSR), with $n_0 = n(\omega_0)$ being the group index at frequency $\omega_0$, and $L$ representing the circumference of the ring. 
        
        Suppose that the ring resonator is modulated by an EOM, introducing a time-dependent perturbation $\delta\epsilon(r, t)$ to the static dielectric function $\epsilon_s(\mathbf{r})$. This perturbation takes the form $\delta\epsilon(r,t) = \delta(r)$cos$(\Omega t+\phi)$, where $\delta(r)$ is the modulation profile, $\phi$ is the modulation phase, $\Omega$ is modulation frequency. For simplicity, we set $\Omega = \Omega_R$. As a consequence, the polarization current density induced by the dynamic modulation can be written as\cite{yuan2018review}
	    \begin{equation}
	        P = \delta\epsilon E = \frac{\delta(r) E_m a_m}{2} (e^{i\phi} e^{i\omega_{m+1}t} + e^{-i\phi} e^{i\omega_{m-1}t}) 
        \end{equation}	   
        where $E_m$ is the eigenmode field distribution of $m$th mode, $a_m$ denotes its modal amplitudes. We can see the induced polarization will resonantly excite modes $m+1$ and $m-1$ at the same time. Consequently, the evolution of the modal amplitudes $a_m$ is governed by the coupled-mode equation:
        \begin{equation}
         	i\frac{d a_m}{d t} = g e^{i\phi} a_{m-1} + g e^{-i\phi} a_{m+1}
        \end{equation}
        where $g$ is the coupling strength determined by the modulation depth and overlap integral between neighbor modes. This evolution equation can be derived from the TB Hamiltonian:
        \begin{equation}
        	H = g \sum_{m} (e^{i\phi} a^{\dagger}_{m+1} a_m +  e^{-i\phi} a^{\dagger}_m a_{m+1})
        \end{equation}
        where $a_m$ is the annihilation operator for the $m$th resonator mode. This Hamiltonian describes a 1D TB model with the nearest-neighbor coupling in the synthetic frequency dimension, where the coupling incorporates a phase component $\phi$ \cite{yuan2016photonic, yuan2018review}, as shown in right figure of Fig.~5a. Intriguingly, the use of modulation frequencies that are integer multiples of $\Omega_R$ enables the realization of long-range coupling in the synthetic frequency dimension\cite{yuan2018synthetic}. Such long-range couplings are extremely challenging to realize in photonic waveguide and CROW systems, where their interactions are typically limited to nearest-neighbor coupling. Moreover, photonic systems with synthetic dimensions can achieve a continuously tunable space of discrete frequencies, for instance, through the use of straight waveguides modulated by electro-optic effects\cite{qin2018spectrum}.

        \begin{figure}[H]
        	\setlength{\abovecaptionskip}{-15pt}
        	\begin{center}
        		\begin{tabular}{c}
        			\includegraphics[height=15cm]{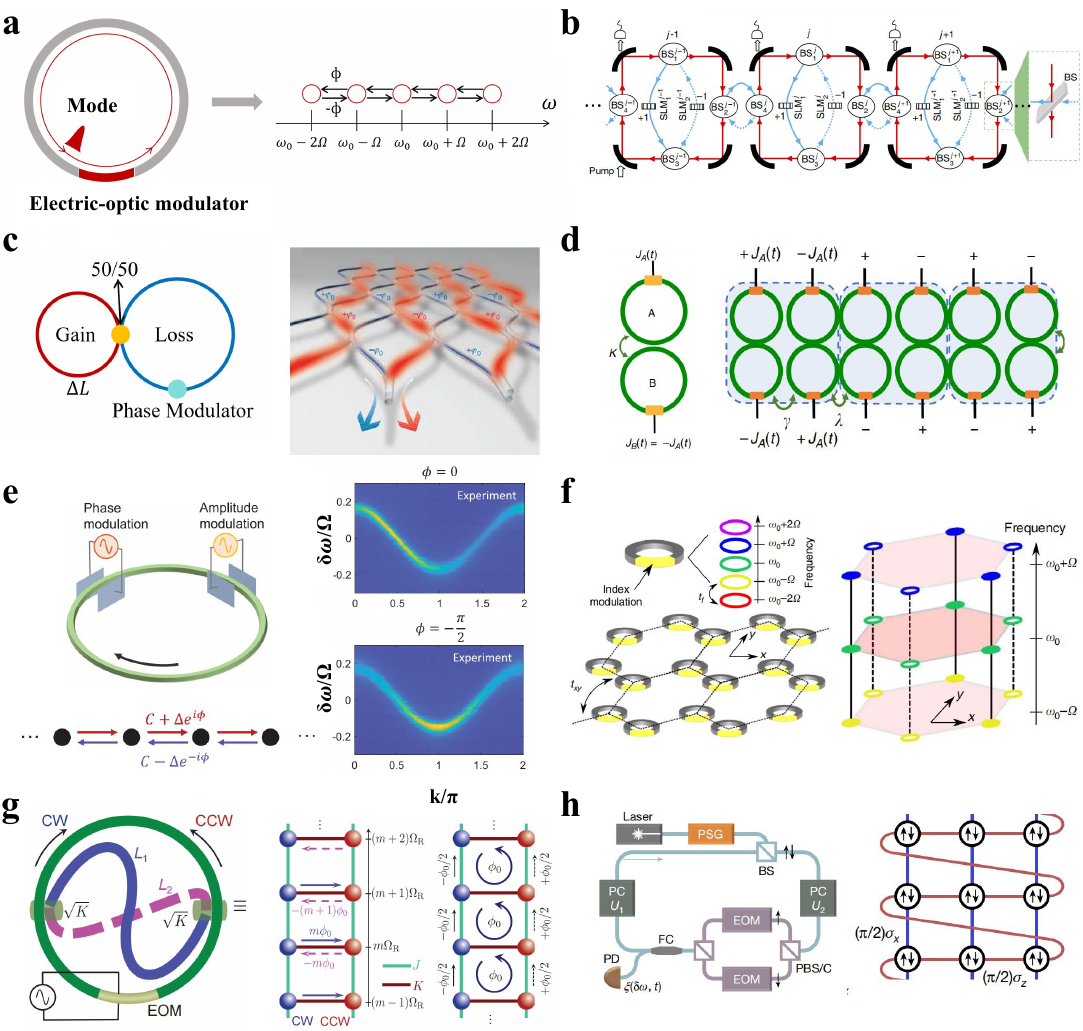}
        		\end{tabular}
        	\end{center}
        	\caption 
        	{Realizations of TB models in photonic systems with synthetic dimension. \textbf{a}. Schematic of a ring resonator with an electric-optic modulator (EOM). The modulation introduces coupling between discrete frequency modes, effectively forming a 1D TB lattice in the frequency dimension. The right panel illustrates the 1D TB model, with modes spaced by the modulation frequency $\Omega$. \textbf{b}. Experimental setup using spatial light modulators (SLMs) and EOMs to couple different orbital angular momentum (OAM) modes. Figure reprinted from Ref.~\cite{luo2015quantum} . \textbf{c}. Gain and loss are introduced in two ring resonators with different lengths to investigate PT symmetry in optical systems. The yellow connection between the two rings represents a 50/50 coupler (left). An equivalent PT-symmetric lattice network (right). Figures adapted from Ref.~\cite{regensburger2012parity} . \textbf{d}. Implementation of the BBH model using coupled ring resonators. Spatial modulation of the coupling strengths ($\gamma$ and $\lambda$) leads to the formation of topologically protected corner states. Figures reprinted from Ref.~\cite{dutt2020higher} . \textbf{e}. A single ring resonator modulated in both phase and amplitude, realizing a non-Hermitian lattice with asymmetric coupling. The panels show the theoretical lattice model (left bottom) and corresponding experimental results for different modulation phases (right). Figures reprinted from Ref.~\cite{wang2021generating} . \textbf{f}. Schematic of a 2D honeycomb array of ring resonators with index modulation (left), which gives rise to a 3D synthetic lattice hosting Weyl points (right). Figures reprinted from Ref.~\cite{zhang2017generation} . \textbf{g}. The left panel shows the schematic of a ring resonator with CW and CCW modes coupling and EOM. The CW and CCW modes are coupled by a 8-shaped coupler (blue and purple lines). The right panel illustrates the resulting lattice with gauge field. Figures adapted from Ref.~\cite{dutt2020single} . \textbf{h}. Schematic of an experimental platform for realizing a non-Abelian SU(2) gauge field. Polarization-dependent rotations are implemented using programmable polarization controllers and EOM (left). The resulting synthetic lattice exhibits SU(2) gauge field (right). Figures adapted from Ref.~\cite{cheng2025non} .
        	} 
        \end{figure}
        
		\subsection{Various examples}
		Synthetic dimensions in lattice systems had been explored in the superconducting qubit circuits to engineer exotic lattice models, such as spin and Hubbard models in higher or even fractal dimensions\cite{tsomokos2010using}. Subsequently, O. Boada et al.\cite{boada2012quantum} proposed a strategy to encode the internal degrees of freedom as an additional synthetic dimension, within the context of ultracold atomic gases. Inspired by these works in lattice systems, the first proposal to realize synthetic dimensions in photonic systems was introduced by X. W. Luo et al. \cite{luo2015quantum}. In their approach, the synthetic dimension is constructed by coupling modes with different orbital angular momentum (OAM) using a pair of spatial light modulators (SLMs), enabling the realization of 2D topological physics within a 1D array of optical cavities, as shown in Fig.~5b. Notably, this 1D structure does not need to be large in scale, thus increasing feasible scale of simulation. Intriguingly, photonic OAM modes can also be utilized to realize a periodically driven Floquet topological insulator, photonic gauge potential in one cavity, exceptional points and other physical phenomena\cite{zhou2017dynamically,yuan2019photonic,yang2022topological,yang2023realization}. Furthermore, X. W. Luo et al. \cite{luo2017synthetic} demonstrated the quantum memory and optical filters can be realized by using synthetic OAM lattices in degenerate cavities.   
		
		On the other hand, temporal degree of freedom can also be exploited to create a synthetic photonic lattice, where the evolution of a sequence of pulses is mapped onto the dynamics of a particle moving on a set of discrete lattice sites\cite{regensburger2011photon,regensburger2012parity}. For example, A. Regensburger et al.\cite{regensburger2012parity} experimentally demonstrated PT symmetric lattices using two coupled fiber loops. Specifically, the experimental setup consists of two fiber loops with a length difference $\Delta L$, connected through a 50/50 coupler. To introduce phase modulation, a phase modulator is placed in the longer loop, as illustrated in Fig.~5c. The length difference $\Delta L$ leads to the discretized arrival times for the optical pulses and to a transverse coupling between neighbor time slots. Consequently, this configuration realizes an equivalent synthetic lattice network. By further introducing optical gain and loss in the two loops via optical amplifiers and amplitude modulators,, a PT-symmetric lattice structure is realized in the synthetic temporal dimension\cite{regensburger2012parity, regensburger2013observation}. Moreover, similar photonic systems can be employed to investigate 1D analogs of “\text{Moir\'e}” lattice\cite{yu2023moire}, to demonstrate almost perfect TR symmetry of a pulse sequence\cite{wimmer2018observation}, and to observe photon-photon isentropic compression-expansion effects\cite{marques2023observation}. Remarkably, such time-periodic temporal photonic lattices can realize dc- and ac-driven Landau-Zener tunnelings (LZTs) between periodic Floquet bands. These systems offer the potential to implement fully reconfigurable LZT beam splitter arrangements, which are of significant importance for signal processing applications\cite{wang2023photonic}.
		
		Thus far, it is evident from the preceding discussion that a single resonator equipped with an EOM can realize a 1D TB model in the synthetic frequency dimension. This enables the exploration of topological edge states, such as those of the SSH model, within a single resonator\cite{zhou2017dynamically}. By extending this principle, arrays of resonators with dynamic modulators can implement higher-dimensional TB models--specifically, 2D (3D) synthetic lattices using 1D (2D) ring resonator arrays. For example, by coupling an auxiliary resonator to a ring resonator with an EOM, a 1D Lieb lattice can be effectively constructed in the synthetic space\cite{li2022observation}. Moreover, the 2D Haldane model has been successfully demonstrated using only three dynamically modulated ring resonators\cite{yuan2018synthetic}. Notably, negative coupling can be readily achieved in such synthetic photonic systems due to the flexibility of EOMs. This enables the realization of more exotic topological phases, such as the BBH model composed of ring resonators, as shown in Fig.~5d. By tuning the coupling strength, e.g. adjusting coupling gaps between the nearest-neighbor cavities along the horizontal axis, topological 0D corner states can be observed\cite{dutt2020higher}. Furthermore, by introducing an additional amplitude modulator with transmission factor $1 + \delta_2$sin$(\Omega t+ \phi)$, non-reciprocal coupling can be effectively achieved, enabling the realization of the HN model in the synthetic dimension, as shown in Fig.~5e\cite{wang2021generating}. Finally, as discussed in the subsection on principles, by setting the modulation frequency range \(\Omega\) to integer multiples of the free spectral range, i.e., $\Omega=m \Omega_R$, it becomes possible to realize arbitrary long-range couplings\cite{wang2021generating, pellerin2024wave}. 
		
	    It is well known that realizing a 3D TB model in conventional photonic systems is challenging. However, many physical phenomena are unique to higher-dimensional systems and have no counterparts in lower dimensions. Photonic systems with synthetic dimensions offer a promising platform to explore such higher-dimensional physics. For instance, Ref.~\cite{lin2016photonic} presents a 2D honeycomb array of ring resonators, each incorporating a phase modulator. The system thus is described by a 3D TB model in the frequency axis, as shown in Fig.~5f. Subsequently, the Weyl-point physics emerges by tuning the modulation phases. Furthermore, by breaking inversion symmetry or introducing artificial gauge fields, the system can host anomalous topological edge states and Fermi arcs\cite{zhang2017generation}. Surprisingly, synthetic dimensions can even enable the realization of 4D lattice that cannot be realized in other photonic systems. For example, A. Dutt et al.\cite{dutt2020higher} constructed a hexadecapole insulator using synthetic frequency dimension, where each unit cell consists of eight site rings arranged in top and bottom layers. Another remarkable case involves a 4D TB model comprising a 3D resonator lattice extended by a synthetic frequency axis, enabling the exploration of the 4D quantum Hall effect\cite{ozawa2016synthetic}. Notably, the idea of synthetic dimensions finds important applications in the realm of optical frequency comb. For example, U. A. Javid et al.\cite{javid2023chip} recently demonstrated a quantum-correlated synthetic crystal, realized through a coherently controlled broadband quantum frequency comb produced in a chip-scale, dynamically modulated lithium niobate microresonator.  
		
		Having discussed the dimensional extension enabled by incorporating a single synthetic dimension, we now turn to the utilization of multiple internal degrees of freedom as synthetic dimensions. For example, A. Dutt et al.\cite{dutt2020single} demonstrated a ring resonator system with two independent physical dimension: a synthetic frequency dimension supported by an EOM, and a synthetic pseudospin dimension formed by the CW and CCW modes at the same frequency, as illustrated in Fig.~5g. Here, CW and CCW modes are coupled via an 8-shaped coupler. As a result, the system creates a synthetic Hall ladder along the frequency and pseudospin degrees of freedom for photons propagating in the ring. Within this platform, a variety of rich physical phenomena have been experimentally observed, including effective spin-orbit coupling, synthetic magnetic fields, spin-momentum locking, a Meissner-to-vortex phase transition, and signatures of topological chiral one-way edge currents. Furthermore, by considering polarization degrees of freedom ($s$ and $p$ polarizations) as pseudo-spin states (spin-up and spin-down) of photons, it becomes possible to construct an SU(2) non-Abelian lattice gauge field within the synthetic frequency dimension\cite{cheng2023artificial}. This is achieved by incorporating two EOMs and two polarization rotators into a ring resonator, as demonstrated in Ref.~\cite{cheng2023artificial}. The experimental realization of such a non-Abelian synthetic lattice is illustrated in Fig.~5h. In this setup, the modulation frequencies of the two EOMs are set to $\Omega_R$ and $m\Omega_R$, respectively, thereby realizing a 2D TB model with with twisted boundary conditions. Importantly, the modulation-induced hoppings in the synthetic dimension depends on the polarization, as indicated by the black arrows in Fig.5h. Consequently, the implementation of non-Abelian physics through the interplay of EOMs and polarization rotators, with the system's output monitored via a photodetector\cite{cheng2025non}. Simultaneously, ring resonators incorporating synthetic dimensions have emerged as a robust platform for investigating dynamics, including topological holographic quench dynamics\cite{yu2021topological}, measurement of dynamic band structure\cite{li2021dynamic}, Bloch transport in the presence of non-linearity and dissipation\cite{englebert2023bloch}, and topological spin pump analogous to Laughlin’s configuration\cite{suh2024photonic}.   
		   		
		\section{Confined-Mie resonance photonic crystals}
		As a significant class of photonic systems for controlling the flow of light, PCs play a pivotal role in the realization of TB models and the exploration of associated topological phenomena \cite{ozawa2019topological,xie2018photonics}. Remarkably, the incorporation of topological physics into PC design has led to promising solutions for practical limitations in photonic applications, such as notable loss and limited bandwidth in current beamformers\cite{wang2024chip}. In this section, we provide a brief overview of the principles of PCs and introduce a recently proposed structure known as CMR-PCs, whose band structures exhibit a precise correspondence with those calculated by TB models with nearest-neighbor couplings.
		
		\subsection{Principle of photonic crystals}
		In addition to the photonic systems discussed in previous sections, PCs composed of artificially structured materials offer an effective platform for realizing TB models. To date, most studies on topological PCs have focused on structures based on periodically arranged dielectric rods or arrays of air holes patterned into dielectric media.\cite{ozawa2019topological, khanikaev2017two, xie2018photonics}. A natural question then arises: how can PCs be employed to realize TB models? To address this, it is essential to begin with Maxwell's equations, as the harmonic modes in PCs are governed by the following master equation\cite{joannopoulos2008molding}:	
		\begin{equation}
			\nabla \times \left( \frac{1}{\varepsilon(\mathbf{r})} \nabla \times \mathbf{H}(\mathbf{r}) \right) = \left( \frac{\omega}{c} \right)^2 \mathbf{H}(\mathbf{r}).
		\end{equation}
		Here, $\varepsilon(\mathbf{r})$ denotes the relative permittivity, $\mathbf{H}(\mathbf{r})$ represents the mode profile of the magnetic field and $\omega$ is modal frequency. Furthermore, this equation can be reformulated as a standard eigenvalue problem by introducing a linear operator $\hat{\Theta}$, defined as 
		\begin{equation}
			\hat{\Theta} \mathbf{H}(\mathbf{r}) = \left( \frac{\omega}{c} \right)^2 \mathbf{H}(\mathbf{r}).
		\end{equation}
		For a given dielectric structure $\varepsilon(\mathbf{r})$, we can solve this equation to find the modes $\mathbf{H}(\mathbf{r})$ and corresponding frequencies $\omega$. Consequently, if one were to construct a crystal consisting of macroscopic uniform dielectric "atoms", the photons in this crystal could be described in terms of a band structure, as in the case of electrons\cite{joannopoulos1997photonic}. At this stage, a fundamental similarity with the TB models becomes evident: both involve solving an eigenvalue equation, where the eigenmodes and eigenvalues correspond to physical observables such as field distributions and modal frequencies, respectively. 
		
		In PCs, the lattice sites are replaced by macroscopic media with different dielectric constants, while the couplings between sites are represented by the spatial overlap of the electromagnetic eigenmodes supported by neighbor dielectric elements \cite{joannopoulos2008molding}. Much more interesting is the regime where these EM eigenmodes (Mie resonances), analogous to electronic orbitals, can hybridize to form bonding and antibonding modes between neighbor dielectric elements\cite{antonoyiannakis1999electromagnetic}. As a result, the coupling strength can be tuned by adjusting the positions of dielectric rods (or air holes), or by modifying the dielectric permittivities of materials\cite{xie2019visualization, yang2018visualization}. However, a particularly significant and attractive difference between PCs and TB models lies in the former's inherent ability to host the orbital degree of freedom. For example, an orbital version of the SSH model has been experimentally realized in a 1D lattice of polariton micropillars\cite{st2017lasing}, demonstrating the practical relevance of orbital degrees of freedom in photonic systems.   
						 
		\subsection{Two-dimensional confined Mie resonance photonic crystals}
		The study of PCs has a influential history, originating from the proposals in 1987 that introduced the concept of photonic bandgaps\cite{yablonovitch1987inhibited, john1987strong}. Since then, PCs have attracted sustained interest due to their ability to prohibit the propagation of electromagnetic waves within specific frequency ranges, enabling a wide range of photonic applications, including PC waveguides\cite{vlasov2005active}, PC fibers\cite{dudley2006supercontinuum}, and PC lasers\cite{park2004electrically}. However, the non-negligible energy losses resulting from fabrication imperfection restricts the further development of photonic devices\cite{ozawa2019topological}. In this context, the topological theories in TB models have emerged as a promising solution to address these challenges.
		
		In a natural parallel with the first discovered insulating topological phase of magnetized electrons, the quantum Hall photonic topological insulator was the first to be theoretically proposed by S. Reghu and F. D. M. Haldane\cite{haldane2008possible}. Later that year, Z. Wang et al.\cite{wang2009observation} experimentally demonstrated robust unidirectional edge states within the photonic bandgap at microwave frequencies. This was achieved by breaking TR symmetry using magneto-optical PCs based on yttrium iron garnet (YIG). However, extending such a model to the optical domain remains a significant challenge, primarily due to the absence of a large magneto-optical response in the optical domain. To overcome this limitation, one way for realizing topological states is to consider internal degrees of freedom of photons as pseudospins, thereby enabling a photonic analog of the quantum spin Hall effect in 2D systems \cite{wu2015scheme,xie2020higher}. Moreover, the proposed configuration is composed of subwavelength dielectric structures, which significantly enhances its experimental feasibility. Another strategy involves the use of quantum valley Hall PCs, which support propagating edge states with weak topological protection. When the inversion symmetry is broken, a complete bandgap appears at the Dirac points. By integrating the Berry curvature around each Dirac point, a topological invariant known as the valley Chern number can be defined. Notably, the zigzag interface between two valley PCs with opposite valley Chern numbers supports edge states\cite{dong2017valley}. Remarkably, these topological PCs hold great promise for a wide range of modern optical applications, such as continuously tunable power splitters\cite{skirlo2014multimode}, pseudospin-polarized\cite{chen2017multiple} and valley-polarized power splitters\cite{chen2018tunable}. They also offer potential breakthroughs in the electrically pumped topological laser\cite{zeng2020electrically} and next generation of terahertz communication technologies\cite{yang2020terahertz,shalaev2019robust,wang2024chip}.

	    Additionally, PCs with higher-order topology have been proposed and realized in kagome\cite{li2020higher}, square\cite{xie2019visualization, chen2019direct} and honeycomb lattices\cite{xie2020higher}. By tuning the distance between neighbor dielectric rods within the same unit cell, the intracell and intercell coupling strengths can be effectively modulated. Notably, both topological edge states and corner states have been observed using near-field scanning technique\cite{xie2019visualization}. Moreover, it has been reported that a higher-order topological PC is fabricated into GaAs slabs with quantum dots embedded, where the coupling between quantum dots and topological corner states can enhance both the photoluminescence intensity and the emission rate\cite{xie2020cavity}. Beyond these periodical cases, other exotic TB models can also be implemented in PCs, including disclination lattices\cite{liu2021bulk} and moiré lattices\cite{oudich2024engineered}. Importantly, 0D topological states in PCs can be exploited in the realization of nanocavity for topological lasers, owing to their spatial localization and robustness against disorder. These properties enable the development of low-threshold topological nanolasers\cite{zhang2020low}. 
	    
		Intriguingly, these physical phenomena can be predicted by analytically solving the Hamiltonian of TB models\cite{ozawa2019topological,gao2020dirac,xie2021higher}. Nevertheless, dielectric PCs exhibit certain fundamental differences from their TB model counterparts. For example, at lower frequencies, the host medium in dielectric PCs supports propagating modes for all frequencies, resulting in a linear dispersion near the Brillouin zone center\cite{khanikaev2017two,ozawa2019topological}. Moreover, the Mie resonance modes of dielectric rod are not exponentially localized, but instead decay as $1/r$ when $r\to \infty$\cite{lidorikis1998tight}. These discrepancies in the band structures lead to deviations from the zero-energy characteristics typically associated with higher-order topological states and topological bound states in the continuum (BICs) in TB models\cite{xiao2024revealing, wang2021quantum,cerjan2020observation}. Furthermore, the high-frequency bands of dielectric PCs are entangled, making it difficult to investigate higher-orbital physics. Fortunately, a recent study\cite{li2024disentangled} has introduced a novel photonic crystal (PC) structure known as the "CMR-PC", whose band structures align closely with that of a TB model with nearest-neighbor couplings. Specifically, by embedding perfect electric conductors (PECs) among the dielectric rods--which can be practically realized using metallic materials at microwave frequencies--the slowly decaying Mie resonance modes are effectively confined, as shown in Fig.~6a. This structure significantly suppresses the long-range interactions, ensuring that the coupling between the dielectric rods is predominantly restricted to nearest neighbors. As a result, the system's band structure closely mimics that of a TB model with nearest-neighbor couplings, faithfully preserving the characteristic chiral symmetry of the latter\cite{asboth2016short} (see right diagram of each CMR-PC structure in Fig.~6a). Therefore, a wide variety of 2D TB models with nearest-neighbor coupling can be perfectly realized in PCs. For instance, authors exhibited the realization of TB models with $C_{3v}$, $C_{4v}$ and $C_{6v}$-symmetries using CMR-PCs\cite{li2024disentangled}. Unlike conventional topological PCs, where the coupling strength is typically tuned by adjusting the positions of dielectric rods\cite{li2020higher,xie2019visualization,chen2019direct}, the CMR-PCs enable coupling control through the variation of the sizes of the embedded PECs. Consequently, the corner and edge states can be realized by increasing the radius of metallic rods at the center and edges, while decreasing it at the corners. Notably, this approach not only simplifies the structure design but also facilitates the realization of interlayer coupling along the $z$-direction. Furthermore, the resulting band structures become disentangled and exhibit clear degeneracy points even at high frequencies, offering new opportunities for exploring higher-orbital physics in PCs, such as topological disclination modes within higher-orbital band gaps\cite{zhang2025hybrid} (see Fig.~6b). Crucially, a series of photonic valley waveguides based on topological higher-orbital disclination states have been constructed and interconnected via distinct topological disclination cores, resulting in robust transmission of hybrid-orbital edge modes\cite{zhang2025hybrid} (see Fig.~6c). In parallel, CMR-PCs have also enabled the precise realization of flat-band physics, as exemplified by the photonic Lieb lattice illustrated in Fig.~6d. To overcome the negative effects of long-range interactions, PECs are strategically inserted between the dielectric rods, enabling the realization of an ideal flat band that closely matches the predictions of a TB model. Moreover, this platform supports the the on-demand design of diverse localized mode patterns, offering great potential for applications in cavity-based photonic devices\cite{li2025realization}.  
		
		\begin{figure}[H]
			\setlength{\abovecaptionskip}{-15pt}
			\begin{center}
				\begin{tabular}{c}
					\includegraphics[height=6.2cm]{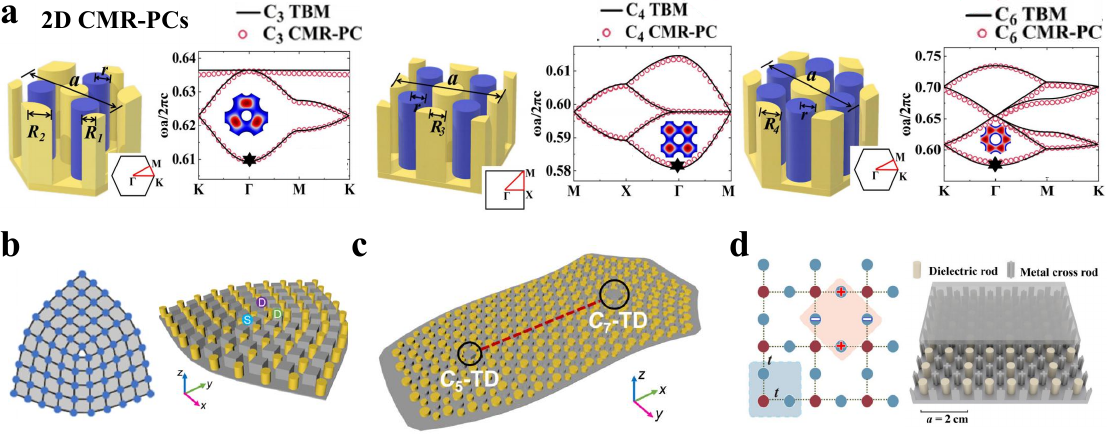}
				\end{tabular}
			\end{center}
			\caption 
			{Realizations of 2D TB-PCs. \textbf{a}. Schematics of 2D $C_{3v}$ (left), $C_{4v}$ (center), and $C_{6v}$ (right) CMR-PCs. Yellow and blue rods represent metallic and dielectric components, respectively. $a$, $r$, and $R_1$ ($R_2$, $R_3$) denote the lattice constant, the radius of dielectric rods, and the radius of metallic rods, respectively. The corresponding band structures compare the results of TB models (black lines) and CMR-PCs (red circles). Insets show the field distributions of the lowest-frequency eigenmodes (black stars). Figures reprinted from Ref.~\cite{li2024disentangled} . \textbf{b}. Schematics of TB model with disclinations (left) and it photonic realization using CMR-PCs (right). \textbf{c}. Schematic of photonic waveguides with topological disclination states. Figures reprinted from Ref.~\cite{zhang2025hybrid} . \textbf{d}. TB model and band structure of the Lieb lattice. Red and blue denote minority (A) and majority (B) sublattices, respectively. The blue square marks the unit cell; the red diamond highlights the compact localized state (left). Realized in CMR-PCs (right). Figures reprinted from Ref.~\cite{li2025realization} .
			} 
		\end{figure}	
		
		\subsection{Three-dimensional confined Mie resonance photonic crystals}
		Most studies on Mie resonance PCs have predominantly focused on 2D systems. This focus primarily stems from the fact that, in 2D PCs, the transverse electric (TE) and transverse magnetic (TM) modes can be separately mapped to scalar wave equations in the mirror symmetry plane\cite{yang2023scalar}. Consequently, the associated topological phenomena can be effectively predicted using TB models. However, the vectorial nature of EM waves in 3D Mie resonance PCs leads to complicated band dispersions, thereby introducing fundamental differences in their band structures relative to those described by TB models\cite{slobozhanyuk2017three,yang2019realization}. Moreover, at lower frequencies, the band structure in the vicinity of the $\Gamma$ point still exhibits linear dispersion, thereby precluding the presence of chiral symmetry in these bands\cite{li2024disentangled}. Furthermore, in 3D PCs, complete photonic bandgaps are relatively rare and typically arise only in specially engineered structures, such as woodpile-like architectures\cite{noda2000full}. To achieve a complete bandgap, it is necessary for the gap to smother throughout the entire 3D Brillouin zone, rather than along a particular plane or high-symmetry line\cite{joannopoulos2008molding}.  
				
		Another strategy involves extending the 2D CMR-PC structure discussed in the previous subsection into the third dimension. This is achieved by inserting metallic plates perforated with air holes between adjacent 2D layers, thereby realizing effective coupling along the $z-$direction, as shown in Fig.~7a\cite{li2024disentangled}. Since only the geometrical parameters of the metallic elements are modified, this configuration ensures efficient coupling between dielectric rods in adjacent layers.	Remarkably, the band structures of 3D CMR-PC exhibits excellent agreement with that predicted by the TB model (see right panel of Fig.7a). This precise correspondence enables the direct prediction of the 3D PC band structures and associated topological phenomena by solving the simple TB model, thereby significantly reducing the computational cost associated with full-wave simulations. Moreover, by tuning the intracell and intercell coupling strengths in the 3D CMR-PC--specifically adjusting the height of the metallic plates $h$ and $h_1$ and the sizes of the air holes $R_1$ and $R_2$--it becomes possible to readily achieve a complete 3D photonic bandgap without the need for complex 3D architectures. Recently, Z. Y. Wang et al.\cite{wang2025realization} experimentally realized a 3D higher-order topological insulator using 3D CMR-PC structures, in which perforated copper plates (golden color) with air holes were adopted to induce vertical interlayer and intralayer couplings, while perforated air foams (white color) were used to fix the metallic and dielectric rods, as illustrated in Fig.~7b. By modulating radii of the metallic rods and the distances between the dielectric rods and the central metallic rods, the 2D surface, 1D hinge and 0D corner topological states were successfully observed, matching well with the TB results. 
		
		Furthermore, 3D CMR-PC structures offer new possibilities for exploring the Dirac hierarchy in the PCs. Y. X. Zhang et al.\cite{zhang2024dirac} proposed that the Dirac hierarchy can be split into lower dimensions in an orderly manner by carefully adjusting the in-plane and out-of-plane coupling strengths in 3D CMR honeycomb-SSH PCs, as shown in Fig.~7c. Through constructing $C_n$-symmetric topological crystalline insulators or breaking mirror symmetry, 1D hinge modes and 0D corner modes can be achieved. Moreover, by implementing an alternating gain-loss configuration in both $y$ and $z$ directions and modulating in-plane and vertical hopping in the 3D CMR-PC (see Fig.~7d), robust non-Hermitian hinge and corner modes at sample edges have been theoretically realized\cite{zhang2024topological}. Beyond the realm of higher-order topological insulator, 3D CMR-PCs successfully open a new for exploiting topological lattice defects, such as 3D topological Dirac-vortex modes. Previously, constructing 3D topological photonic structures with Kekulé distortion was challenging due to fundamental differences between 3D TB models and PCs. Remarkably, B. Yan et al.\cite{yan2025topological} theoretically propose and experimentally demonstrate the existence of topological Dirac-vortex modes in 3D CMR-PCs by inducing Kekulé-distortion. In this structure, the Kekulé-distortion are implemented by displacing the dielectric rods from their original positions, while interlayer couplings are introduced via air holes perforated in the copper plates, as illustrated Fig.~7e. As a result, topological Dirac-vortex modes, which are bound to and propagate vertically along 1D vortex line defects, are successfully observed. More Notably, the use of metallic plates perforated with air holes has not only facilitated the construction of 3D CMR-PC, but has also enabled the realization of a variety of exotic phases, including 3D PC semimetals\cite{wang2022higher}, 3D Chern insulators\cite{liu2022topological}, and 3D axion insulators\cite{liu2025photonic}. In addition to CMR-PCs, another structural platform has been explored to realize TB behavior in 3D photonic systems. For instance, B. Yang et al.\cite{yang2023scalar} proposed a nested meta-crystal composed of interconnected coaxial waveguides, which effectively mimics scalar-wave-like dispersions and matches TB model predictions (see Fig.~7f). These designs further broadening the landscape of 3D topological photonics.
		
		\begin{figure}[H]
			\setlength{\abovecaptionskip}{-15pt}
			\begin{center}
				\begin{tabular}{c}
					\includegraphics[height=10cm]{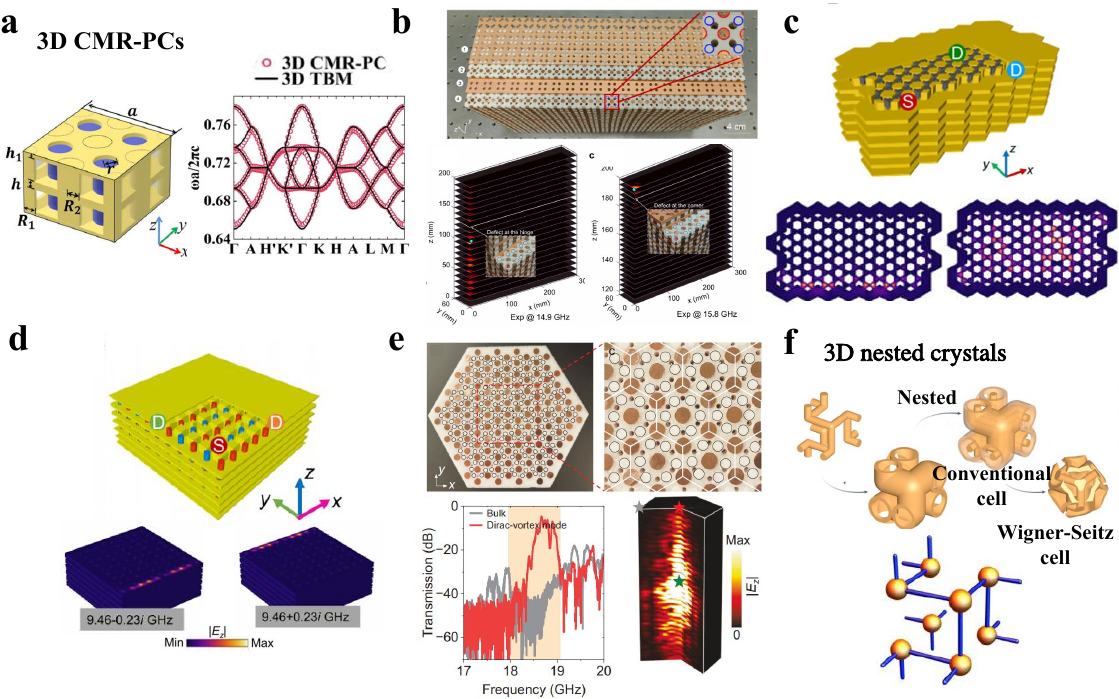}
				\end{tabular}
			\end{center}
			\caption 
		{Realizations of 3D TB-PCs. \textbf{a}. Schematic of a 3D CMR-PC, heights of the interlayer metal plates are $h$ and $h_1$ and radii of metallic rods are $R_1$ and $R_2$. The right panel illustrates the striking similarity in the band structures between the 3D TB model (where the intracell coupling is equal to the intercell coupling) and the 3D CMR-PC (where $h=2h_1$ and $R_1=R_2$) band structure. Figures reproduced from Ref.~\cite{li2024disentangled} . \textbf{b}. Experimental realization of 3D CMR-PCs, and the zoom-in view of the unit cell, consisting of two perforated copper plate layers and two air foam layers embedded by metallic (red circles) and dielectric (blue circles) rods (top). Observation of 1D topological hinge states and 0D topological corner states (bottom). Figures reprinted from Ref.~\cite{wang2025realization} . \textbf{c}. 3D schematic of a finite-size CMR-PC with a honeycomb lattice (top). Bottom panels show the field distributions of hinge modes (left) and surface modes (right). Figures adapted from Ref.~\cite{zhang2024dirac} . \textbf{d}. 3D CMR-PC formed by non-Hermitian stacked topological layers. Red and blue rods indicate dielectric rods with gain and loss, respectively (top). Normalized electric field distributions of the non-Hermitian structure (bottom). Figures adapted from Ref.~\cite{zhang2024topological} . \textbf{e}. Photograph of the fabricated 3D CMR-PC with Kekulé distortion (top). Measured transmission spectra of bulk (gray) and topological Dirac-vortex modes (red) are shown at bottom left. Electric field distribution of the Dirac-vortex mode excited by a point source (green star) is displayed at bottom right. Figures reprinted from Ref.~\cite{yan2025topological} . \textbf{f}. 3D nested crystals. This panel illustrates a TB model (bottom) constructed from nested meta-crystals of connected coaxial waveguides (top). Figures reprinted from Ref.~\cite{yang2023scalar} .
	       } 
        \end{figure}

		\section{Conclusion and perspective}
        The TB model serves as a representative and analytically tractable theoretical model in condensed matter physics\cite{kittel2018introduction}. It provides a concise yet powerful tool for understanding band dispersion, orbital hybridization, gauge fields and topological phases\cite{harrison2012electronic,asboth2016short}. Extending the TB model to photonic systems not only facilitates the direct emulation of physical phenomena in electronic systems, but more crucially, offers a new paradigm for constructing, controlling, and inversely designing electromagnetic wave propagation. Moreover, by establishing a direct correspondence between structural parameters and the Hamiltonian, TB photonics provides a systematic theoretical foundation for the development of robust and multifunctional photonic devices. For example, the SSH and Haldane models have been successfully implemented in optical platforms\cite{ozawa2019topological}, demonstrating that topologically protected states can significantly enhance immunity to defects and fabrication imperfections.  
        
        In recent years, the research focus on TB photonics has shifted from geometric structure design to machine learning-assisted optimization, from single-mode systems to multimode systems, from 1D and 2D architectures toward three- and even higher-dimensional configurations\cite{ozawa2019topological,liu2022topological,liu2025photonic,li2025symmetry}. This evolution reflects a significant transition in the field -- from feasibility to achieving precision. In this context, realizing an exact mapping between TB models and photonic systems is emerging as a core prerequisite for advancing the next generation of photonics. Intriguingly, the recently proposed CMR-PCs, with their various tunable aspects (as illustrated in Fig.1), are expected to become one of the key platforms for realizing TB photonics. In the following, we discuss several future directions for CMR-PCs from both theoretical and application oriented perspectives.
        
        Theoretically, TB-PCs may provide a favorable platform for introducing gradient fields, such as gradual variations in the position, size, or refractive index of dielectric rods, which could allow simulation of effective gauge field\cite{aidelsburger2018artificial}. Such design, deeply based on the TB models where hopping amplitudes and on-site energies vary, may offer novel routes to emulate complex gauge structures not readily accessible in photonic systems. One other interesting theoretical direction is the construction of 2D and 3D non-Hermitian TB-PCs\cite{zhang2024topological}. So far, experimental realizations of 2D and 3D non-Hermitian PCs are still very limited, especially for realizing TB models that exhibit non-reciprocal couplings. We expect to see PT-symmetric transitions through introducing gain and loss on the dielectric rods and the observation of novel non-Hermitian topological states. 
        
        Another promising direction is to utilize TB-PCs to explore higher-orbital physics. This is enabled by the characteristic high frequency band structure of TB-PCs, which hosts disentangled higher-orbital bands and degeneracy points\cite{li2024disentangled}, thereby allowing for the selective excitation and control of higher-orbital modes. In contrast to fundamental $s$-modes, higher-orbital modes exhibit anisotropic spatial profiles, and their couplings may depend on relative orientation, requiring tensorial rather than scalar descriptions of intercell couplings\cite{medinad}. In addition, the presence of higher-orbital modes may enable the realization of vortex states carrying orbital angular momentum in PCs through TR symmetry breaking\cite{chen2024chiral,carlon2019optically,wang2024ex}. Furthermore, TB-PCs incorporating higher-orbital modes could possibly support higher-order topological phases\cite{zhang2023realization, medinad}. Altogether, these prospects may establish TB-PCs as a powerful platform for studying higher-orbital phenomena, which could reveal new physical effects and topological phases in PCs. 
        
        In parallel, recent developments in Fock-state lattices (FSLs) may provide new perspectives for PC design. In FSLs, photon states are mapped onto discrete lattice sites, with coupling strengths that depend on occupation numbers\cite{yuan2024quantum}. This leads to inhomogeneous couplings, which may induce a topological edge or synthesize a pseudo magnetic field. The resulting eigenmodes are not governed by discrete translation symmetry, but are instead encoded in internal photon numbers, setting them apart from Bloch states. Such lattices have been experimentally realized in superconducting quantum circuits, where phenomena such as pseudo-Landau levels, the valley Hall effect and chiral edge currents have been observed\cite{deng2022observing}. Moreover, FSLs  have also been implemented in phononic crystals, enabling the emergence of 3D SU(3) Landau levels\cite{peng2025ideal}. Crucially, in this system, internal quantum numbers can be directly reconstructed from eigenmode correlations. In this context, the matrix-inspired principle of TB-PC provides precise control over resonance modes and inter-site couplings, making it possible to mimic Hamiltonians analogous to those of FSLs. This may provide new insights for PC design, including the realization of all-band-flat lattice\cite{yang2024realization} and the exploration of many-body physics\cite{yuan2024quantum}.
        
        Moreover, TB-PCs might facilitate the design and exploration of sophisticated lattice geometries by precisely controlling the positions and sizes of dielectric rods and metallic elements. Examples include pyrochlore\cite{ezawa2018higher}, moiré superlattices\cite{yu2023moire}, hyperbolic lattices\cite{kollar2019hyperbolic}, and fractal structures\cite{biesenthal2022fractal}, which are challenging to realize in conventional PCs. These sophisticated lattices, extensively studied in TB models for their exotic band topology and physical phenomena, may unveil new physical insights and novel topological phases when implemented in PCs. Perhaps most importantly, the direct correspondence between TB-PC designs and the matrices that describe TB models could enable theoretical advances beyond band theory. Intriguingly, the exploration of PC designs may possibly be transformed into solving eigenvalues and eigenvectors of the corresponding TB models. Hence, by leveraging mathematical theory such as matrix theory\cite{zhang2011matrix} and spectral theory\cite{arveson2002short}, we expect to see the matrix-inspired photonics.
        
        From the application perspective, TB-PCs may offer a promising pathway toward high-throughput and inverse photonic design\cite{shang2022experimental}. The explicit mapping between structure parameters and the Hamiltonian could allow rapid screening of photonic structures with desired electromagnetic responses. When combined with machine learning, TB-PCs might facilitate the efficient exploration of novel matter phases, functionalities and material configurations. Another particularly important future direction is the realization of all-dielectric TB-PCs without relying on metallic components. Such an approach is critical for minimizing Ohmic losses and might enable the practical development of topological photonic devices operating at optical frequencies\cite{yang2020terahertz}. Moreover, compared to conventional 3D PCs with complete 3D band gaps\cite{joannopoulos2008molding}, TB-PCs exhibit greater structural simplicity and fabrication accessibility. Hence, they may provide a robust platform for realizing essential building blocks in optical logic gates and multifunctional photonic chips.  
        
        In summary, TB photonics provides a clear theoretical framework for understanding and constructing topological phases in the photonic systems, while simultaneously fostering a new paradigm of matrix-inspired PC design. Looking ahead, TB-PCs are poised to enable key breakthroughs in the realization of sophisticated TB models, novel topological states, and the inverse design of photonic devices. It is foreseeable that TB photonics will serve as a vital bridge between fundamental research and practical applications, advancing photonics toward a new era of enhanced controllability and intelligent functionality.		
		
		\section{Acknowledgments}
		B. Y. Xie acknowledges fundings from the National Natural Science Foundation of China under grants No. 62475225 and No. 12404187, the National Key R\&D Program of China under grant No. 2023YFA1407700, GuangDong Basic and Applied Basic Research Foundation under grant No. 2024A1515012031, Shenzhen Science and Technology Innovation Commission under grant No. CYJ20240813113619025, Stable Support Program for Higher Education Institutions of Shenzhen under grant No. 20220817185604001. M. H. Lu acknowledges fundings from National Key R\&D Program of China under grant No. 2023YFA1406904. P. Zhan acknowledges fundings from the National Natural Science Foundation of China under grants No. 12174189 , and the National Key R\&D Program of China under grant No. 2023YFA1406901. X. L. Zhuo acknowledges support from Department of Science and Technology of Guangdong Province (grant Nos. 2023A1515110091, 2023QN10C200).

		
		\bibliography{report}   
		\bibliographystyle{spiejour}   
		

		
		
	\end{spacing}
\end{document}